%% file: DengNeuronalCircuitModel.tex
\documentclass[12pt]{article}
\usepackage[dvips]{color}
\usepackage{graphicx}
\usepackage{amssymb,amsmath,graphics,times}

\input{definitions}

\begin{document}
\centerline{\Large \sffamily Mechanistic Model to Replace Hodgkin-Huxley Equations}
%\vskip .1in
%\centerline{\Large \sffamily Voltage-Gating, and Excitable Membranes}

%\centerline{\Large \sffamily Mechanistic Neuron Model}

\bigskip
\centerline{\sffamily Bo Deng\footnote{Department of Mathematics,
University of Nebraska-Lincoln, Lincoln, NE 68588. Email: {\tt
bdeng@math.unl.edu}}}

\bigskip
\noindent{\bf Abstract: In this paper we construct a mathematical model for excitable membranes by introducing circuit characteristics for ion pump, ion current activation, and voltage-gating. The model is capable of reestablishing the Nernst resting potentials, all-or-nothing action potentials, absolute refraction, anode break excitation, and spike bursts. We propose to replace the Hodgkin-Huxley model by our model as the basis template for neurons and excitable membranes. }

\bigskip
\noindent{\bf 1. Introduction.} When Hodgkin and Huxley constructed their model for the squid giant axon (\cite{hodgkin1952quantitative}) they were fully aware of their model's drawbacks because it was only a phenomenological fit to their experimental data. They commented specifically that different empirical forms should fit the same data or even better. They were right and other researchers saw the same problem too \cite{agin1963some,cole1968membranes,agin1972excitability}. Alternative models were concocted (\cite{agin1963some}) but never gained any traction because there was no point to replace one {\it ad hoc} model by another arbitrary one. However, replacing a phenomenological model by a mechanistic one is a different matter entirely.

By mechanistic it is meant for a model to have as few hypotheses as possible that apply to physical processes or objects of a same type. Newton's inverse-distance-squared law for gravitation is the first and one perfect example of mechanistic modeling because it applies to all macroscopic bodies of mass. Goldman's derivation (\cite{goldman1943potential}) of Nernst potentials across cell membranes is mechanistic because it applies to all ion species. In contrast Hodgkin and Huxley's individual treatments of the sodium and the potassium currents are not mechanistic because their hypothesis for the sodium ion does not apply to the potassium ion or vice versa. Researchers must have tried but failed because other than various variations of the HH model no mechanistic model can be found in the literature.

The purpose of this paper is to fill this literature gap. The idea is to model the membrane as a circuit of devices each is defined by a current-voltage characteristics. We will model the sodium-potassium ion exchanger pump by the $IV$-characteristics that the time-rate of change of the current is proportional to the power of the pump. We will model the ion channel activation and the voltage-gating by one unified $IV$-characteristics that the voltage-rate of change of the conductance is proportional to the conductance. We will demonstrate that the resulting conductance-adaptation model is capable of reproducing all known phenomena of the HH model and much more, and hence provides a mechanistic alternative to the HH model and a model template for other types of excitable membranes in general.

\bigskip
\noindent{\bf 2. The Result.} The approach to modeling the squid giant axon as an electrical circuit began with Cole's work (\cite{cole1932surface,cole1940permeability,cole1949dynamic}) that the axon membrane had a rather narrow range around 1$\mu$F per square centimeter for membrane capacitance. Let $V$ be the intracellular potential difference (inside potential minus outside potential) and $I=CdV/dt$ denote the capacitance current. Let $\Kpcurrent,\ \Napcurrent$ denote respectively the potassium and the sodium ion current using the outward direction as the reference direction. Rather than the leakage current considered by the HH model we consider instead the voltage-gating current $\Gpcurrent$ measured outward. Last let $\Ecurrent$ be the net remaining current with the inward reference direction. Then the following equation must be the first equation of any circuit model of the membrane by Kirchhoff's Current Law
\[
C\frac{dV}{dt}=-[\Kpcurrent+\Napcurrent+\Gpcurrent-\Ecurrent].
\]
The aim of this section is to derive the functional forms for the potassium, the sodium, and the gating currents based on our proposed mathematical models for ion pump, ion channel activation, and voltage-gating, and to obtain the resulting model for the membrane.

\medskip\noindent{\em Ion Pump.} Let $q$ denote the intracellular charge difference (charge concentration inside minus outside) of a given ion species across the cell membrane.
The following as a mathematical model for the pump was proposed in \cite{deng2009conceptual}
\begin{equation}\label{mdPumpModel}
\frac{d^2q}{dt^2}=\phi \frac{dq}{dt} q
\end{equation}
with $\phi$ a parameter. If we let $I=dq/dt$ be the current through the ion pump and $V=q/C$ be the across-membrane potential over the pump with $C$ being the membrane capacitance. Then the model can be interpreted as the $IV$-characteristics of the pump as an electrical device because the model is equivalent to
\[
\frac{dI}{dt}=\lambda IV
\]
with $\lambda=C\phi$ referred to as the pump parameter in the unit of per time per
voltage. As the product $IV$ represents power. The pump characteristics can be stated as the change of the ion pump current is proportional to the power of the pump. This is a way to model the energy transfer of the pump when ATP is converted to ADP in exchange for ion transportation across the membrane.

%%%%%%%%%%%%%%%%%%%%%%%%%%%%%%%%%%%%%%%%%%%%%%%%%%%%%%%%%%%%%
%%%%%%%%%%%%%%%%%%%%%%%%%%%%%%%%%%%%%%%%%%%%%%%%%%%%%%%%%%%%%
%%%%%%%%%%%%%%%%%%%%%%%%%%%%%%%%%%%%%%%%%%%%%%%%%%%%%%%%%%%%%
\begin{figure}[t!]

\centerline
{\scalebox{.8}{\includegraphics{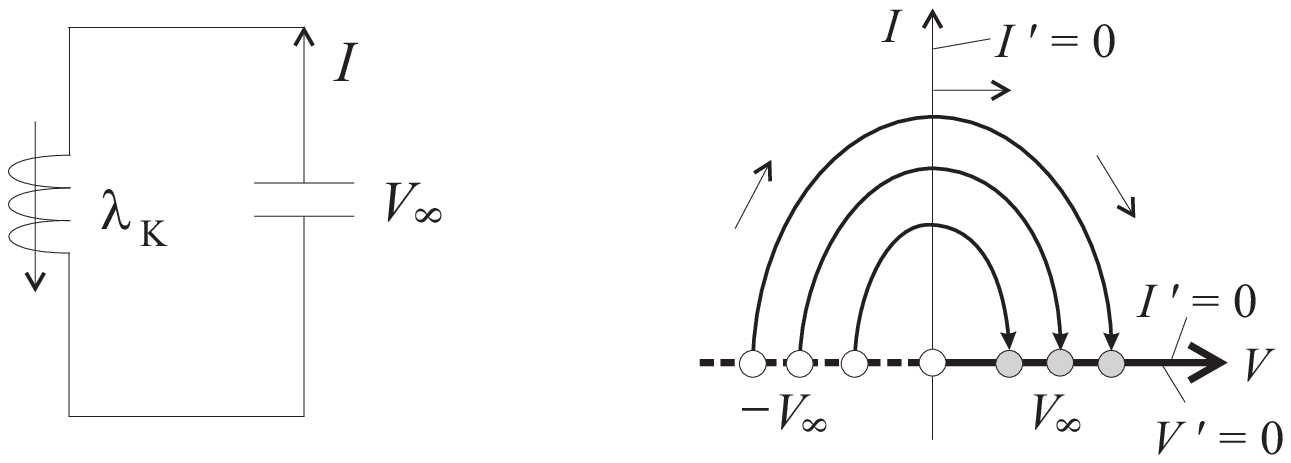}}}
\vskip .07in
\centerline{(a)\hskip 2.5in (b)$\qquad$}

\vskip .1in
{ \centerline
{\scalebox{.8}{\includegraphics{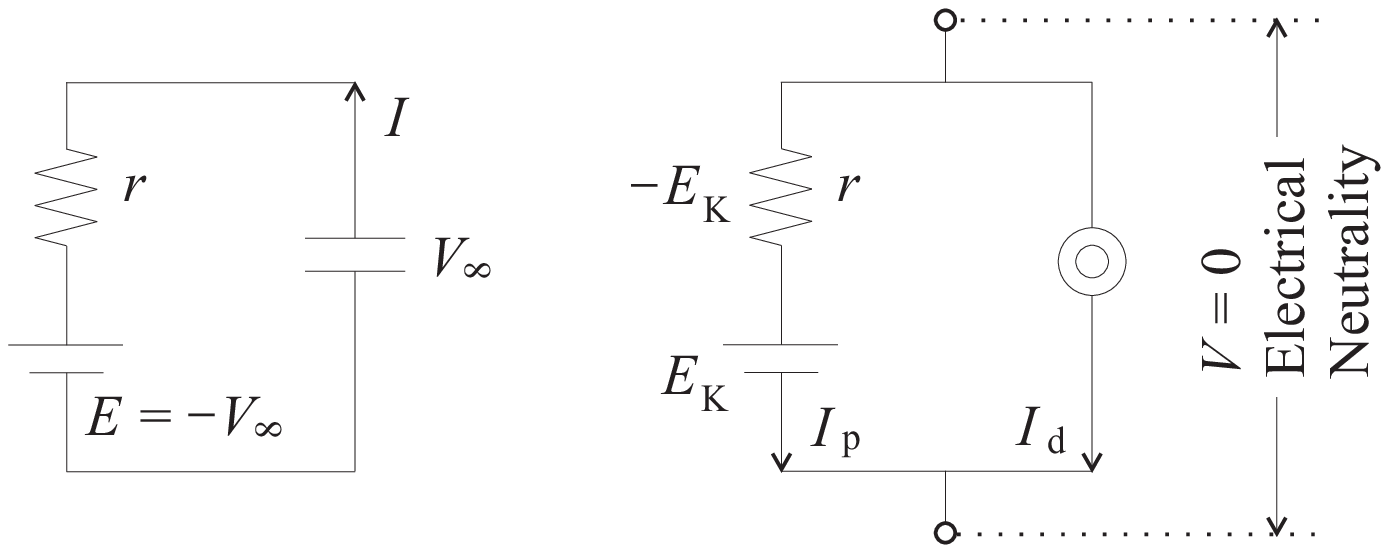}}}%samplefigure.eps}}}
}
\vskip .07in
\centerline{(c)\hskip 2.5in (d)$\qquad$}

\caption{(a) Circuit diagram of the potassium ion pump with
the capacitive membrane. (b) The phase plane
portrait of the circuit equations (\ref{eqPumpEquation}) with
$\gamma=0$. (c) A $CRE$-circuit approximation of the pump circuit. (d) The Nernst potential under electrical neutrality and across-membrane diffusion.}\label{figIonPumpDynamics}
\end{figure}
%%%%%%%%%%%%%%%%%%%%%%%%%%%%%%%%%%%%%%%%%%%%%%%%%%%%%%%%%%%%%
%%%%%%%%%%%%%%%%%%%%%%%%%%%%%%%%%%%%%%%%%%%%%%%%%%%%%%%%%%%%%

The minimal circuit consisting a pump, a resistor, and a capacitor in loop can be understood completely, c.f. Fig.\ref{figIonPumpDynamics}. The ordinary differential equations for the circuit is
\begin{equation}\label{eqPumpEquation}
CV'=I, \qquad I'=-\lambda I(V+\gamma I).
\end{equation}
Eliminating the time variable the equation becomes $dI/dV=-\lambda C (V+\gamma I)$ which being a linear equation can be solved explicitly. In fact all important insights can be obtained by dropping the resister term ($\gamma=0$), for which the solution with the initial condition $V(0)=0,I(0)=I_0\ge 0$ is
\begin{equation}\label{eqPumpDynamics}
V(t)=V_\infty\frac{1-e^{-\lambda V_\infty t}}{1+e^{-\lambda
V_\infty t}} \quad\hbox{and}\quad  I(t)=2\lambda CV_\infty^2\frac{e^{-\lambda V_\infty t}}{[1+e^{-\lambda V_\infty t}]^2}
\end{equation}
with
\[
V_\infty=\sqrt{\frac{2I_0}{\lambda C}}.
\]
The solution for the current means the pump is always unidirectional. That is, if it is positive at one point, say $I_0>0$, then it stays positive all the time. Equally important, as $t\to -\infty$, $\lim V(t)=-V_\infty,\ \lim I(t)=0$ and as $t\to +\infty$, $\lim V(t)=V_\infty,\ \lim I(t)=0$.
This means if the membrane is negatively charged in the past ($\sim -V_\infty<0$) then the membrane will be positively charged eventually ($\sim V_\infty>0$) if the pump's transporting direction is from outside to inside as with the case of the potassium pump. Similarly, for the sodium pump, it will transport all sodium ion from inside to outside in asymptote.

We demonstrate next that the minimal pump-capacitor circuit is exponentially close to a resistor-battery-capacitor circuit in loop with the battery value given by $E=-V_\infty$, c.f. Fig.\ref{figIonPumpDynamics}. The differential equation for the circuit is: $CV'=I=-(V+E)/r$ with the resistance $r$ to be determined. Solving it with the zero voltage initial condition
$V(0)=0$ as we did for the pump circuit we have
\[
V(t)=-E[1-e^{-t/(rC)}]=V_\infty[1-e^{-t/(rC)}].
\]
So if we choose the resistance as
\begin{equation}\label{eqBatResistance}
r=\frac{1}{\lambda C V_\infty}
\end{equation}
then the dynamics for the battery circuit and the pump circuit are exponentially close for large $t$ since the pump dynamics is approximately $V(t)=V_\infty[1-e^{-\lambda V_\infty t}]/[1+e^{-\lambda V_\infty t}]\sim V_\infty[1-e^{-t/(rC)}]$, converging to the same resting potential $V_\infty$ at the same exponential rate $e^{-\lambda V_\infty t}$, and starting from
the same initial value $V(0)=0$. That is, the ion pump can be approximated by a
conductor-battery in series with the conductance $1/r=\lambda CV_\infty$. This approximation becomes  more accurate as time goes by. This gives a mechanistic justification for Hodgkin and Huxley's modeling of ion channel currents by battery-driven ion channels.

\medskip\noindent{\em Nernst Potential.} The pump dynamics above says that in the absence of other forces such as diffusion a particular ion pump will eventually deposit all ions of the same type from one side of the membrane to the other side. We determine next the value of the asymptotic potential $V_\infty$ of the pump under the influence of diffusion and the electrical neutrality condition of the membrane, c.f. Fig.\ref{figIonPumpDynamics}, which is an illustration for the potassium ion. By Bernstein's hypothesis (\cite{bernstein1912electrobiologie}) this inward current due to the electromotive force is balanced out by the outward diffusion assuming the membrane
is permeable to potassium ions. Specifically, let $I_p,I_d$ denote the pump
current and the diffusive current as shown. Let the reference
direction be inward and be defined by the across-membrane variable $x$ from outside to
inside and $a$ be the thickness of the membrane. Let $n(x)$ be the ion density at position $x$ and let $E_{\rm K}=-V_\infty$ be the equilibrium state reached when these two currents cancel each other out under the neutrality assumption $V=0$. Then the same Bernstein-Goldman equation below must hold (\cite{goldman1943potential,hodgkin1949effect})
\[
0=I_d+I_p=D\left(-\frac{dn}{dx}\right)+\mu n \left(-\frac{E_{\rm
K}}{a}\right),
\]
with $D$ the diffusion constant and $\mu$ the mobility parameter (\cite{hodgkin1949effect}). Solving this linear equation in $n$ from $x=0$ to $x=a$, and then expressing the result in $E_{\rm
K}$ to obtain the following which is exactly the potassium ion's Nernst resting potential
\[
E_{\rm K}=-\frac{D}{\mu}\ln\frac{[{\rm K}]_i}{[{\rm K}]_o}
\]
with $[{\rm K}]_i=n(a),\ [{\rm K}]_o=n(0)$ denoting the inside and outside potassium concentration, respectively. The same derivation can be used to obtain the sodium ion's Nernst potential.

\medskip\noindent{\em Ion Channel Activation.}
With the presence of various types of ions and charged molecules, the across-membrane potential $V$ is the manifestation of their aggregate. The Independence Hypothesis in electrophysiology holds that the opening and closing of an ion gate is a function of the voltage not of the other ions. As a consequence individual ion currents are modeled by the ohmic $IV$-characteristics form $I=g(V-E)$. By eliminating the spatial effect of axon by their voltage-clamp experiments Hodgkin and Huxley showed (\cite{hodgkin1952measurement,hodgkin1952currents,hodgkin1952components,hodgkin1952dual}) the conductance $g$ undergo changes throughout the course of an action potential and had the insight to propose it to be a function of the voltage, that is, $g=g(V)$. Their work showed that throughout the course of an action potential the ion conductances for sodium and potassium ions, $\Naconductance, \Kconductance$, increase from near zero when the membrane is at rest and then decrease to near zero again when the action potential ends, hence implying that there are ion channels which open to increase their conductances and close to decrease them. Here is where the HH model becomes empirical rather than mechanistic because of its use of arbitrary functional forms for the sodium and potassium channel conductances.

\begin{figure}%[ht]

\centerline{
\parbox[l]{2.5in}{
\centerline
{\scalebox{.5}{\includegraphics{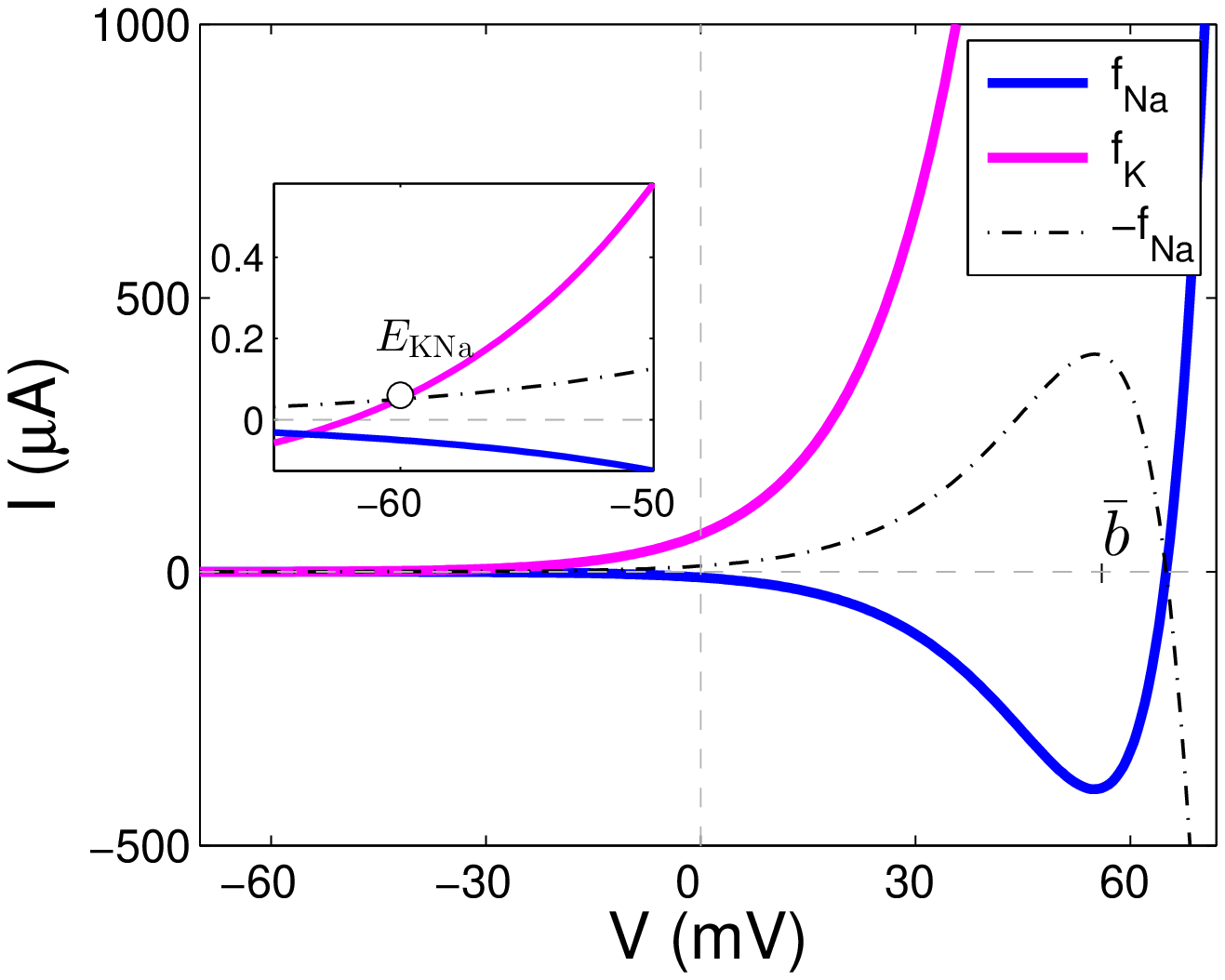}}}
}
\hskip 2cm
\parbox[l]{2.5in}{
\centerline
{\scalebox{.5}{\includegraphics{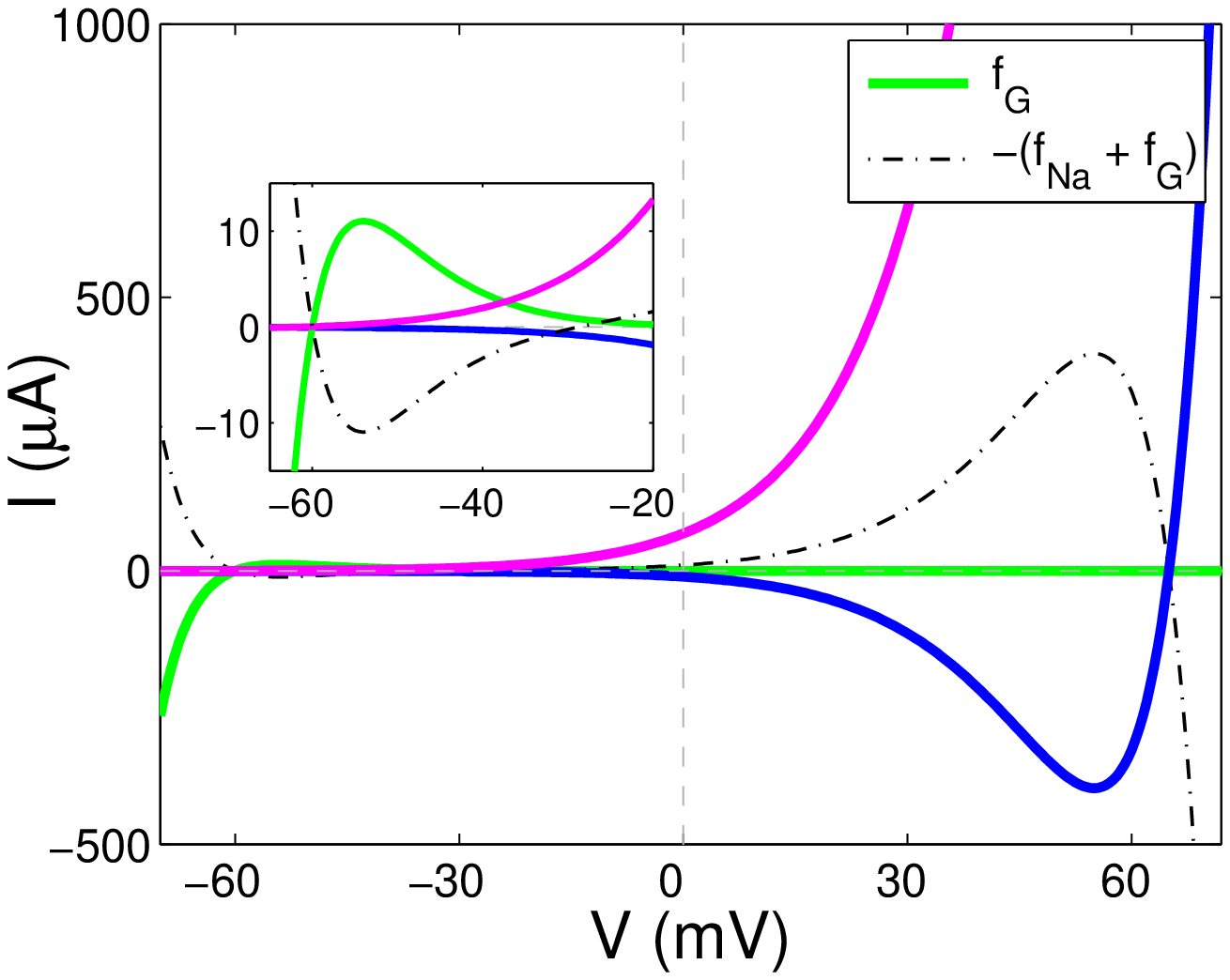}}}
}
}
\vskip .07in
\centerline{\ \ \ \ (a)\hskip 3.16in (b)}

\caption{(a) Typical $IV$-characteristics curves for the sodium,  potassium, $\NaIVf,\KIVf$, respectively. The left-most intersection of $I=\KIVf$ and $I=-\NaIVf(V)$, $E_{\rm KNa}$, is a first approximation of the membrane resting potential $E_m$. But the geometric configuration of the two $IV$-curves implies the KNa-equilibrium state $E_{\rm KNa}$ is almost always unstable. Here $\Naabattery-\Nabconstant$ is the minimum point of the $\NaIVf$-curve. (b) A similar plot as (a) except for the addition of the gating $IV$-curve $\GIVf$ with $\Gabattery=E_{\rm KNa}$. The true resting membrane potential $E_m$ is the left-most intersection of $I=\KIVf(V)$ and $I=-(\NaIVf(V)+\GIVf(V))$, as shown in the inset. %The %same plot as (a) but for various values of parameter $\Nabconstant$, showing approximated $E_m$ %increases in $\Nabconstant$.
}\label{figDModelIVCurves}
\end{figure}

By channel activation we mean channel opening and closing. We propose the following model for both types of activation
\begin{equation}\label{mdActivationModel}
\Delta g\sim g\Delta V\qquad \hbox{ equivalently }\qquad \frac{dg}{dV}=\frac{g}{b}
\end{equation}
with $b>0$ the activation parameter in the dimensional unit of voltage. This means $V$ is increasing ($\Delta V>0$) if and only if the conductance $g$ is increasing ($\Delta g>0$). That is, the changing of conductance (opening or closing of ion channels) with respect to depolarizing or hyperpolarizing potential is proportional to the conductance. It is a double-edged sword --- positive feedback with increasing voltage and negative feedback with the opposite. Solving it we get the exponential activation law $g(V)=ae^{V/b}$. Normalizing it at the ion's Nernst potential with $g(E)=\bar g$ we get
\[
g(V)=\bar ge^{(V-E)/b}.
\]
Namely, $a=\bar ge^{-E/b}$. The parameter $g(E)=\bar g$ is referred to as the intrinsic conductance. It is a simple exercise to check that the corresponding $IV$-curve $I=f(V)=\bar ge^{(V-E)/b}(V-E)$ has a unique minimal point at $\bar b=E-b$ whose value is $-\bar g be^{-1}$, c.f. Fig.\ref{figDModelIVCurves}. To the right side of $\bar b$, the $IV$-curve $f$ is increasing, crossing the $V$ axis at the only resting potential $E$. To the left side of the minimum criticality $\bar b$, $f$ is decreasing, giving rise to a varying negative conductance. It approaches $I=0$ asymptotically from below as $V$ approaches negative infinity. For reasons which will become clearer later we will refer to parameter $b$ as the activation range parameter, measuring the minimal current point from the Nernst potential.

The proposed activation characteristics has this property of dichotomy. If the membrane voltage is increasing in time, $dV/dt>0$, then the activation gates open up exponentially fast in conductance as $dg/dt=dg/dV\cdot dV/dt=kg$ with $k=(dV/dt)/b>0$. In contrast, if the membrane voltage is decreasing in time, $dV/dt<0$, then the activation gates close down exponentially fast as well as $dg/dt=kg$ with $k=(dV/dt)/b<0$. It is a model for both opening activation and closing activation with the same voltage-specific exponential rate. That is, opening gates beget more gates opened and closing gates triggers more gates closed.

For the potassium and sodium $IV$-characteristics we get respectively
\[
\KIVf(V)=\Kbarconductance e^{(V-\Kabattery)/\Kbconstant}(V-\Kabattery), \qquad
\NaIVf(V)=\Nabarconductance e^{(V-\Naabattery)/\Nabconstant}(V-\Naabattery).
\]
In the operating range of action potentials between the potassium and sodium Nernst potentials, $(\Kabattery,\Naabattery)$, only the critical potential $\Naabattery-\Nabconstant$ of the sodium current may lie, predicting that the potassium current is always outward and the sodium current is always inward which in turn has two different phases: increased entry into the cell to the left of the critical point $\Naabattery-\Nabconstant$ and decreased entry to the right of the criticality. That is, inside this operating range the sodium and potassium $IV$-curves behave differently qualitatively, the former usually have a negative conductance branch but the latter's conductance is always positive.

%%%%%%%%%%%%%%%%%%%%%%%%%%%%%%%%%%%%%%%%%%%%%%%%%%%%%%%%%%%%%
%%%%%%%%%%%%%%%%%%%%%%%%%%%%%%%%%%%%%%%%%%%%%%%%%%%%%%%%%%%%%
%%%%%%%%%%%%%%%%%%%%%%%%%%%%%%%%%%%%%%%%%%%%%%%%%%%%%%%%%%%%%
%%%%%%%%%%%%%%%%%%%%%%%%%%%%%%%%%%%%%%%%%%%%%%%%%%%%%%%%%%%%%
%%%%%%%%%%%%%%%%%%%%%%%%%%%%%%%%%%%%%%%%%%%%%%%%%%%%%%%%%%%%%
%%%%%%%%%%%%%%%%%%%%%%%%%%%%%%%%%%%%%%%%%%%%%%%%%%%%%%%%%%%%%

There is a quantitative difference between the intrinsic conductances $\Kbarconductance$ and $\Nabarconductance$. First we know from the derivation of the Nernst potential above (specifically (\ref{eqBatResistance})) that the conductance is
\[
\Kconductance(0)=\Kbarconductance e^{-\Kabattery/\Kbconstant}=\frac{1}{r}={\lambda CV_\infty}
\]
with $V_\infty\sim |\Kabattery|$. There is no need for a precise value of $V_\infty$, an estimate of its range or average of its range suffices. Say $\Kabattery \sim -60$ and $V_\infty\sim 60$. This gives rise to an estimate of
\[
\Kbarconductance\sim {\lambda C 60}e^{\Kabattery/\Kbconstant}={\lambda C 60}e^{-60/\Kbconstant}.
\]
Similarly, if we use $\Naabattery\sim +70$, a similar estimate is obtained for the sodium conductance:
\[
\Nabarconductance\sim {\lambda C 70}e^{\Naabattery/\Nabconstant}={\lambda C 70}e^{70/\Nabconstant}.
\]
The conclusion is if the $b$-parameter values for both ions are comparable, say in the decade range, then relative to each other, the intrinsic potassium conductance $\Kbarconductance$ should be a few orders smaller than the intrinsic sodium conductance $\Nabarconductance$. From the HH model the intrinsic potassium conductance is worked out to be $\Kbarconductance=0.0229$ m.mho per square centimeter. As a result a guessed value is used for the intrinsic sodium conductance $\Nabarconductance=100$ m.mho/cm$^2$, both are used below and for the illustrations of the $IV$-curves in Fig.\ref{figDModelIVCurves}.

\medskip\noindent{\em Voltage-Gating.} A major advance in neurophysiology was made in the 1970s (\cite{armstrong1971interaction}) by the discovery of the voltage-gating phenomenon whereby there is a small pulse-like outward current opposite to the inward entry of sodium ions during the onset of action potential across the squid giant axon membrane when the membrane is depolarizing. The gating current is due to the release of charged molecules from the sodium channel pores in responding to some conformational changes of the pores to the depolarizing voltage. (This gating current was not recognized by Hodgkin and Huxley but was nonetheless captured by their meticulously fitted model.)

For the $IV$-curve of the gating current we propose first it has a resting potential $\Gabattery$, which may be taken to be the intersection of the sodium and potassium $IV$-curves, calling it the provisional resting potential $E_{\rm KNa}$,  and second, it is nonlinear ohmic of a similar form as for channel activation $I=\GIVf(V)=\Gconductance(V)(V-\Gabattery)$  except for a negative proportionality:
\begin{equation}\label{mdGatingModel}
\frac{dg}{dV}=-\frac{g}{b}
\end{equation}
with $b>0$. That is, positive proportionality is for activation and negative proportionality is for gating. Gating conductance recedes in proportion to itself with depolarizing voltage ($\Delta g<0$ if and only if $\Delta V>0$). Solving it the corresponding $IV$-curve becomes
\[
I=\GIVf(V)=\Gbarconductance e^{-(V-\Gabattery)/\Gbconstant}(V-\Gabattery).
\]
For this type of function, it has the unique maximum point at the critical voltage $\Gabattery+\Gbconstant$ right of the gating equilibrium $\Gabattery$.  For $\Gabattery<V<\Gabattery+\Gbconstant$ the gating current is increasing in $V$ and outward, preventing the membrane from depolarization. Its maximum defines the minimum threshold for excitation currents to clear in order for action potentials to generate. For this reason, parameter $\Gbconstant$ is referred to as the gating range parameter, similar to the activation range parameters $\Kbconstant,\ \Nabconstant$.

Notice that once the threshold is cleared and the membrane is depolarizing (below $V=0$ and $dV/dt>0$), the gating conductance drops exponentially fast with $dg/dt=dg/dV\cdot dV/dt=-kg$ with $k=(dV/dt)/b>0$, permitting the generation of an action potential. When the membrane is hyperpolarizing (below $V=0$ and $dV/dt<0$), gating becomes active again with the conductance approaching the intrinsic conductance $\Gbarconductance$. To the left side of the gating reversal potential, $V<\Gabattery$, the gating current $I=\GIVf(V)$ is negative (inward), effectively shutting down hyperpolarization and restoring the membrane to its resting state at the same time.

In addition, the sodium-gating parallel combination $IV$-curve, $I=\NaIVf(V)+\GIVf(V)$, is typically $N$-shaped. The left knee point of the $N$-characteristics is near the gating criticality $\Gabattery+\Gbconstant$ and the right knee point is near the sodium criticality $\Naabattery-\Nabconstant$. This is because the tail part of each curve beyond its critical point is exponentially flat. Furthermore, the membrane resting potential $\mbattery$ is the intersection of the potassium $IV$-curve $I=\KIVf(V)$ and the reflection of the sodium-gating $IV$-curve $I=-(\NaIVf(V)+\GIVf(V))$ (so that $\KIVf(\mbattery)+\NaIVf(\mbattery)+\GIVf(\mbattery)=0$). It is to the right side of $\Gabattery$ and below $\Naabattery$, and is always stable, see Fig.\ref{figDModelIVCurves}. Negative conductance has always been a theoretical conundrum (\cite{hodgkin1952components,moore1959excitation,fitzhugh1961impulses,yamamoto1965negative,agin1972excitability}). For our model it is a mechanistic consequence to channel activation (\ref{mdActivationModel} and voltage-gating  (\ref{mdGatingModel}).

%%%%%%%%%%%%%%%%%%%%%%%%%%%%%%%%%%%%%%%%%%%%%%%%%%%%%%%%%%%%%
%%%%%%%%%%%%%%%%%%%%%%%%%%%%%%%%%%%%%%%%%%%%%%%%%%%%%%%%%%%%%
%%%%%%%%%%%%%%%%%%%%%%%%%%%%%%%%%%%%%%%%%%%%%%%%%%%%%%%%%%%%%
\begin{figure}%[ht]

\centerline{\scalebox{.7}{\includegraphics{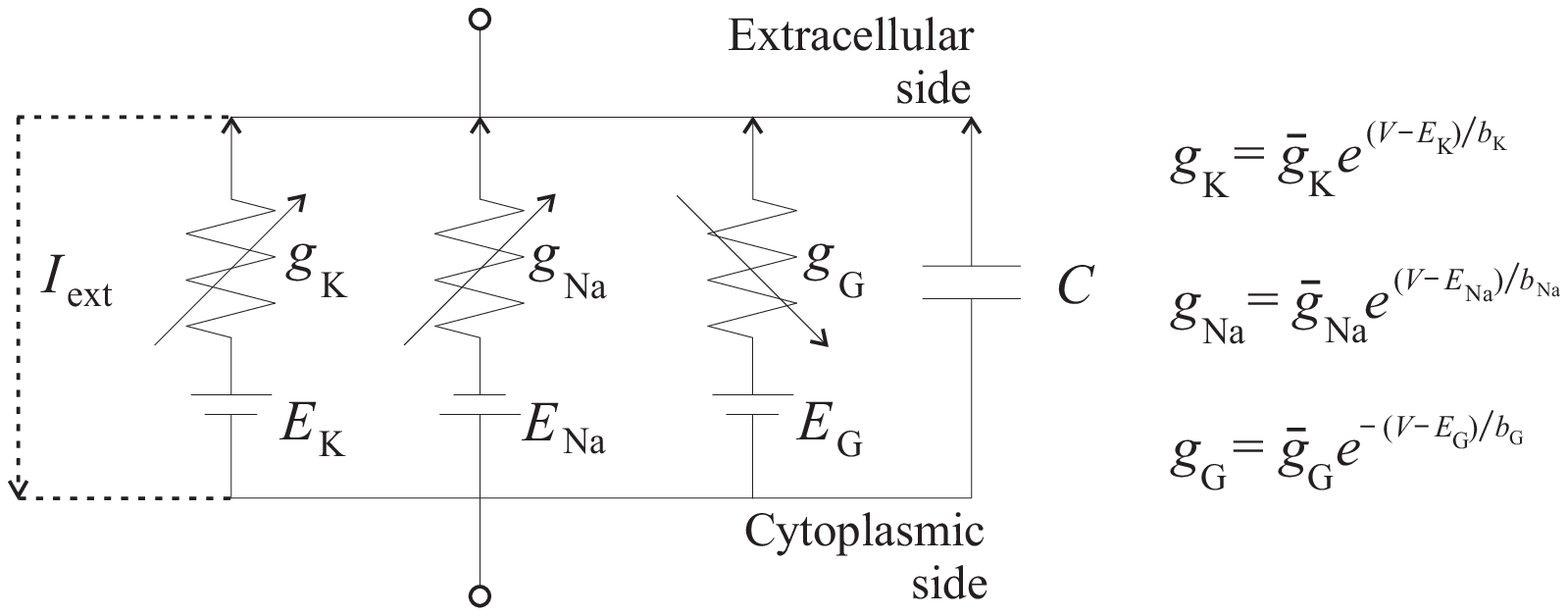}}}

\caption{A circuit representation of model (\ref{eqDengModel}) of which a voltage-gating current replaces the leakage current in the HH model. The activation conductance and the gating conductance are of the same mathematical form but exactly opposite in the direction of the membrane potential.}\label{figDModelCircuit}
\end{figure}
%%%%%%%%%%%%%%%%%%%%%%%%%%%%%%%%%%%%%%%%%%%%%%%%%%%%%%%%%%%%%
%%%%%%%%%%%%%%%%%%%%%%%%%%%%%%%%%%%%%%%%%%%%%%%%%%%%%%%%%%%%%
%%%%%%%%%%%%%%%%%%%%%%%%%%%%%%%%%%%%%%%%%%%%%%%%%%%%%%%%%%%%%

\medskip\noindent{\em Conductance Adaptation and Circuit Model.} Although each current's $IV$-curve is accessible as an equilibrium current for each clamped voltage, in transient the voltage driving conductance $g(V)=\bar ge^{\pm (V-E)/b}$ cannot be realized instantaneously during action potential generation and propagation. There is a time delay or time-course adaptation. We propose the following conductance adaptation following the idea of Hodgkin and Huxley (\cite{hodgkin1952quantitative}):
\begin{equation}\label{mdAdaptationModel}
\frac{dp}{dt}=\tau(e^{\pm (V(t)-E)/b}-p)
\end{equation}
with $\tau$ being a time constant. Thus, instead of the $IV$-curve the corresponding current's time course is given by $I(t)=\bar g p(t)(V(t)-E)$ with $p(t)$ determined by the adaptation equation above.

Putting all these assumptions together we obtain the following mathematical model for the squid giant axon and excitable membranes in general:
\begin{equation}\label{eqDengModel}
\left\{\!\!\!\!\begin{array}{ll} & C\Vvoltageprime=-[\Kbarconductance n (\Vvoltage-\Kabattery)+\Nabarconductance m (\Vvoltage-\Naabattery)+\Gbarconductance h (\Vvoltage-\Gabattery)-\Ecurrent]\\
& {n}' = \KTimeConstant(e^{(\Vvoltage-\Kabattery)/\Kbconstant}-n)\\
& {m}' = \NaGTimeConstant(e^{(\Vvoltage-\Naabattery)/\Nabconstant}-m)\\
& {h}' = \NaGTimeConstant(e^{-(\Vvoltage-\Gabattery)/\Gbconstant}-h)\\
\end{array}\right.
\end{equation}
where $\Ecurrent$ denotes any intracellular current other than the potassium, the sodium, the gating, and the capacitive currents. As for the conductance adaptation time constants, we assume the sodium channel and the gating channel share the same constant $\NaGTimeConstant$ because they are structurally bonded together. The dimensionless variables $n,m,h$ are not the same as of the HH model but are borrowed here to pay tribute to Hodgkin and Huxley's seminal work.

\medskip\noindent{\em Best Fit of Model to Data for Parameter Estimation.} A piece of mathematics only remains as a conceptual model for any physical process unless and until it is best-fitted to the process to fix a parameter point or a parameter range of the model. Otherwise different qualitative behaviors of the conceptual model would forever remain as unrealized and untested possibilities. To this end we re-sampled the action potential data of axon 17 from Fig.12 of \cite{hodgkin1952quantitative} and fitted our model to the data to see how well or badly the model performs.

First we fix some parameter values and exclude them from the best-fit process because they are known. These are: the capacitance $C$, the two ions Nernst resting potentials $\Kabattery,\ \Naabattery$, the intrinsic potassium conductance $\Kbarconductance$. The membrane capacitance is   $C= 1\mu{\rm F}$ per square centimeter. The Nernst potentials are extracted from \cite{hodgkin1952quantitative} which had them shifted up by the Goldman-Hodgkin-Katz potential $E_r$ which was estimated by \cite{hodgkin1949effect} to be around -47.5 mV, leading to $\Kabattery=-59.5$ and $\Naabattery=67.5$ respectively. (As a result the HHAxon17 data is shifted down by the supposedly membrane resting potential $E_r$.) As for the intrinsic potassium conductance $\Kbarconductance$, it is the value of $\Kconductance(\Kabattery)$. As mentioned above it can be worked out to be $\Kbarconductance=0.0229$ m.mho per square centimeter from \cite{hodgkin1952quantitative}.

%%%%%%%%%%%%%%%%%%%%%%%%%%%%%%%%%%%%%%%%%%%%%%%%%%%%%%%%%%%%%
%%%%%%%%%%%%%%%%%%%%%%%%%%%%%%%%%%%%%%%%%%%%%%%%%%%%%%%%%%%%%
%%%%%%%%%%%%%%%%%%%%%%%%%%%%%%%%%%%%%%%%%%%%%%%%%%%%%%%%%%%%%
\begin{figure}%[ht]

\centerline{
\parbox[l]{3in}{
\centerline
{\scalebox{.5}{\includegraphics{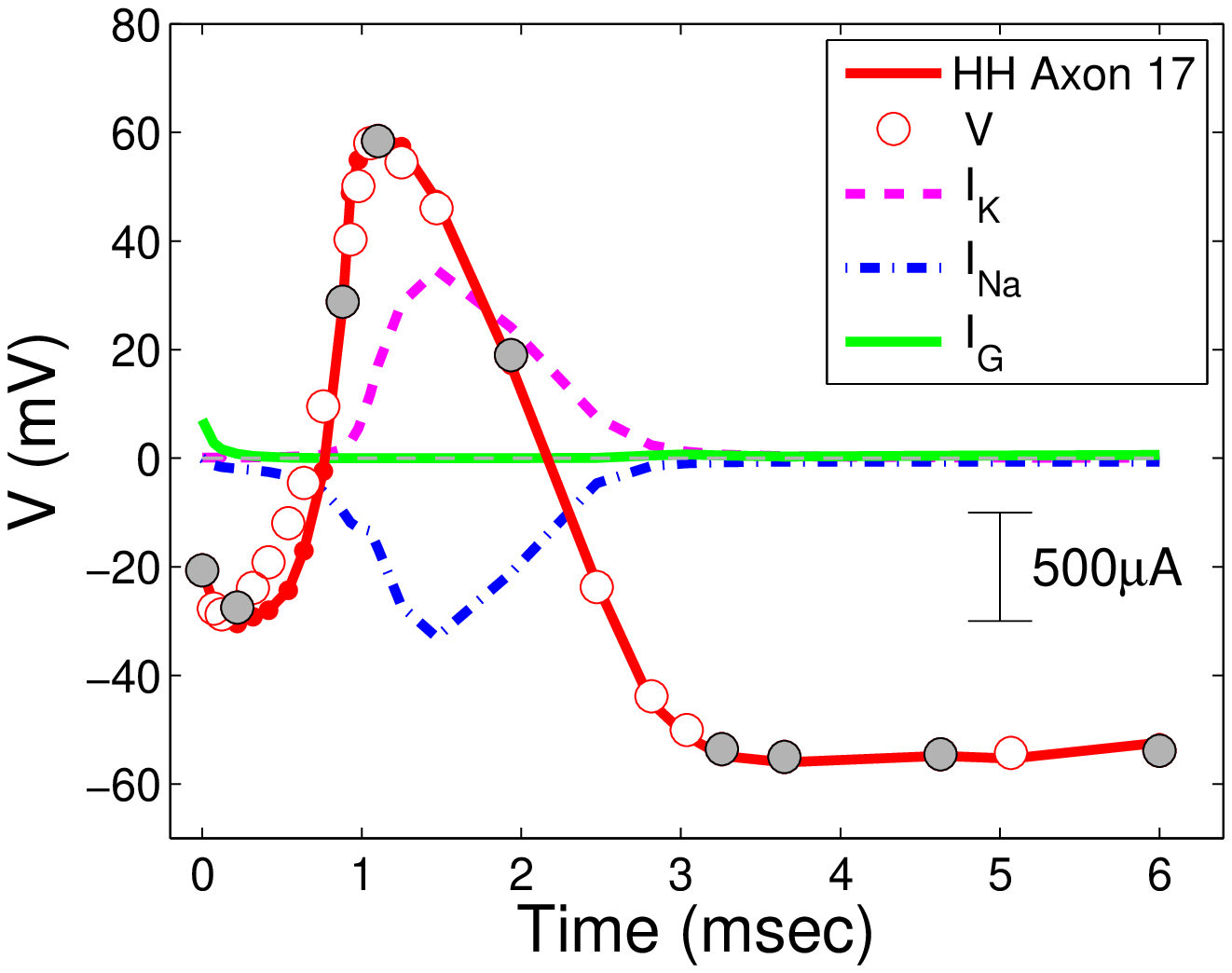}}}
}
\hskip 1cm
\parbox[l]{3in}{
\centerline
{\scalebox{.5}{\includegraphics{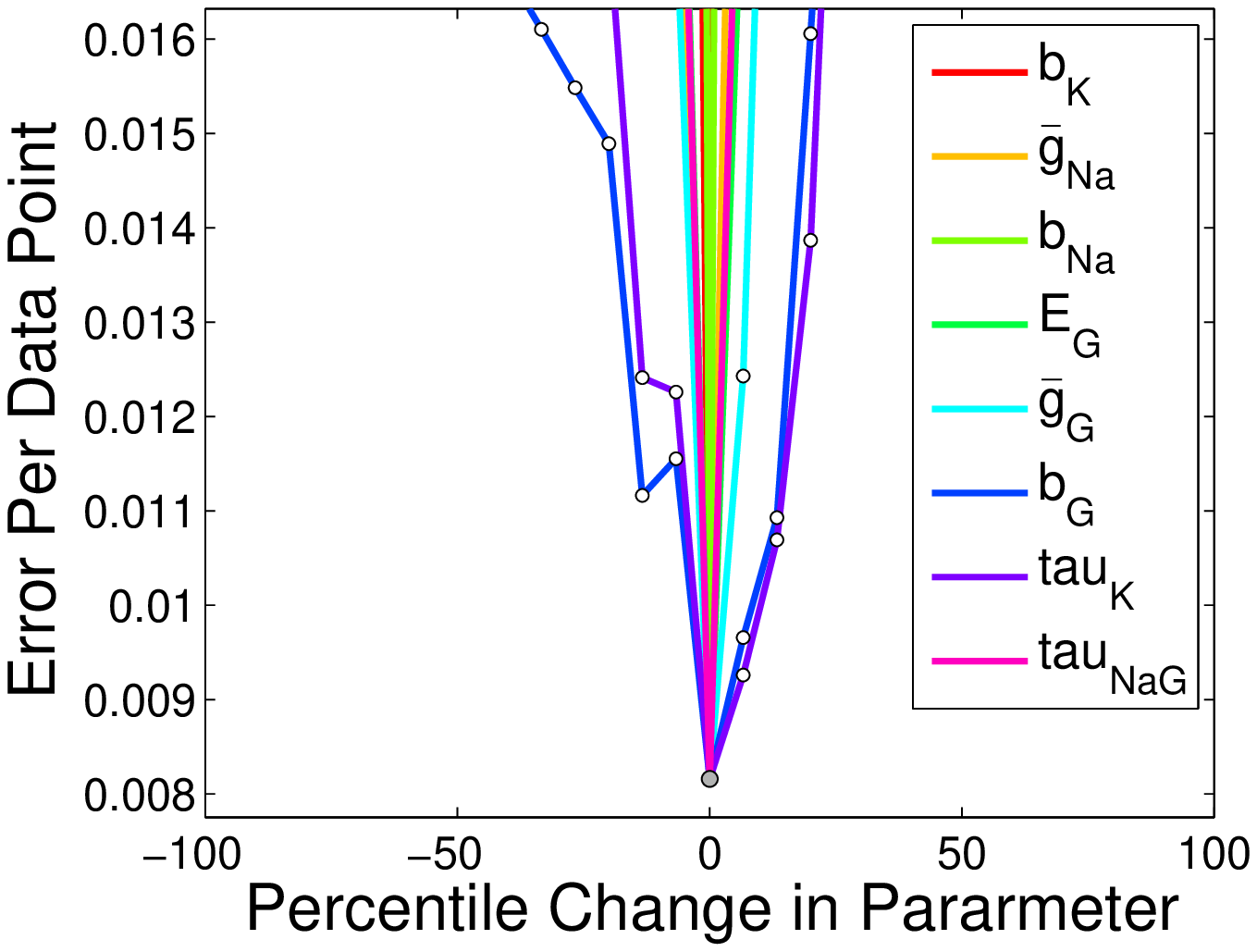}}}
}
}
\vskip .07in
\centerline{\ \ \ \ (a)\hskip 3.3in (b)}

\vskip .1in

\centerline{
\parbox[l]{3in}{
\centerline
{\scalebox{.5}{\includegraphics{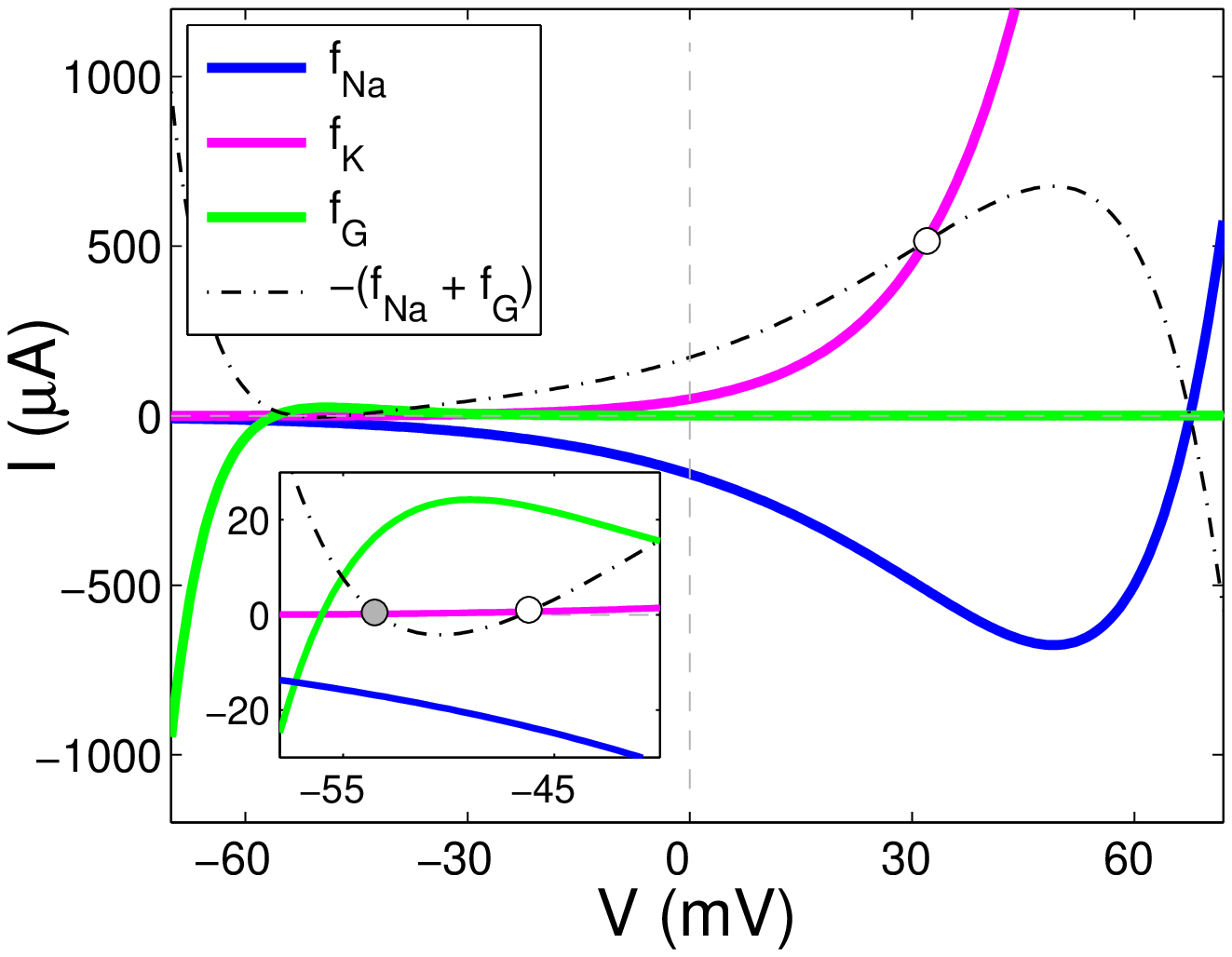}}}
}
\hskip 1cm
\parbox[l]{3in}{
\centerline
{\scalebox{.5}{\includegraphics{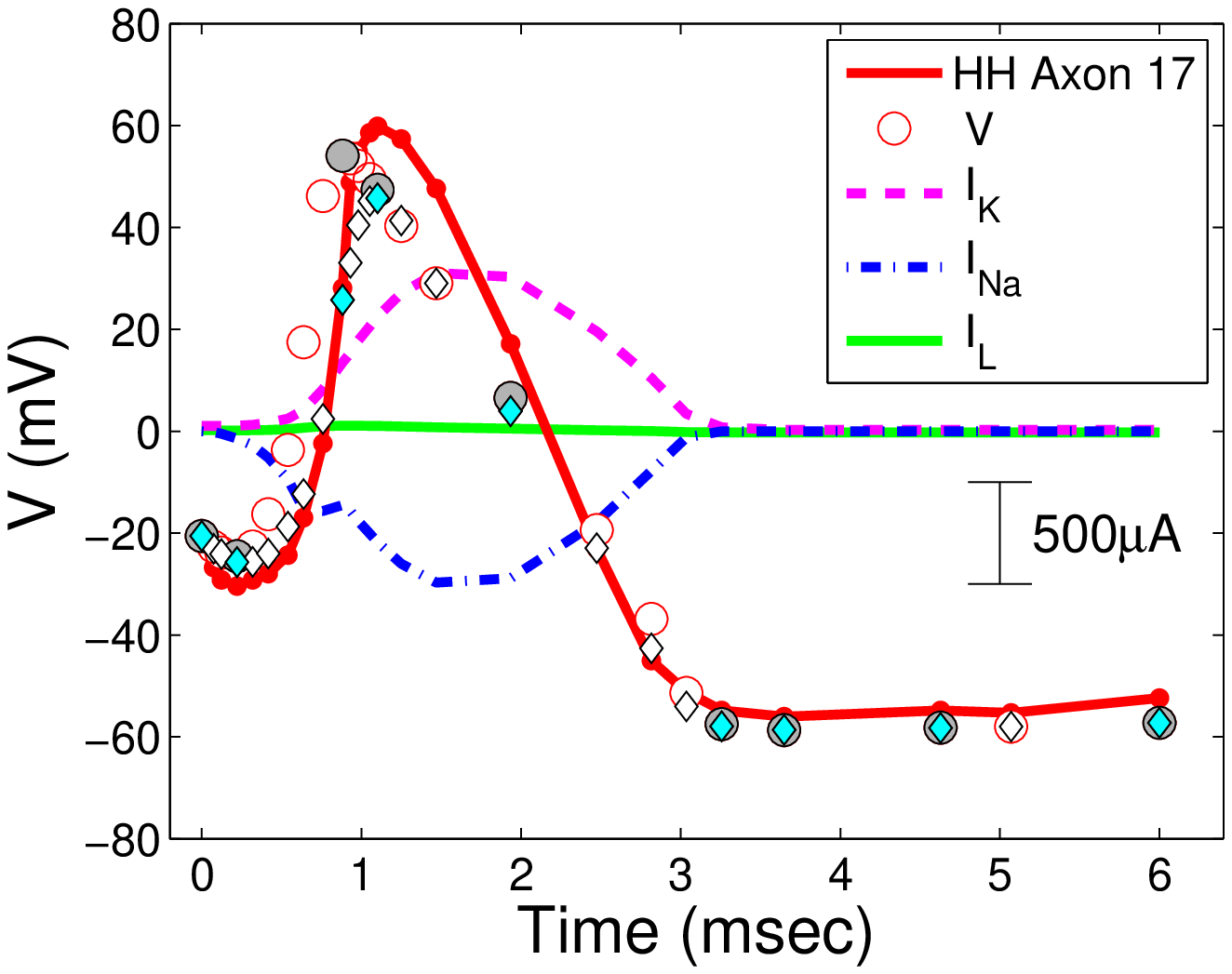}}}
}
}
\vskip .07in
\centerline{\ \ \ \ (c)\hskip 3.3in (d)}

\caption{(a) Following the initial guess from the text, the best-fitted parameter values for the circuit model (\ref{eqDengModel}) are: $\Kabattery=-59.5$, $\Kbarconductance=0.0229$, $\Kbconstant=16.6$, $\Naabattery=67.5$, $\Nabarconductance=100$, $\Nabconstant=18.4$,  $\Gabattery=-56$, $\Gbarconductance=9.3333$, $\Gbconstant=7.0667$, $C=1$, $\KTimeConstant=0.8667$, and $\NaGTimeConstant=10$. Filled data points are used for the second-order error function. (b) The per-data-point error function in each searched parameter centered at the best-fitted point whose error value is 0.0082. (c) $IV$-curves of the best-fitted model, showing three equilibrium states with the left-most being the resting membrane potential $E_m$. (d) A similar plot for the original fit of the HH model to the same data, and a best fit showing the voltage trace only (diamond marker) starting at the original parameter values of the HH model with searching parameters not shared with our model (\ref{eqDengModel}). The original fit error is 0.0595 and the best fit error is 0.0395, both are worse than the best fit of our model.}\label{figBestModelFit}
\end{figure}
%%%%%%%%%%%%%%%%%%%%%%%%%%%%%%%%%%%%%%%%%%%%%%%%%%%%%%%%%%%%%
%%%%%%%%%%%%%%%%%%%%%%%%%%%%%%%%%%%%%%%%%%%%%%%%%%%%%%%%%%%%%
%%%%%%%%%%%%%%%%%%%%%%%%%%%%%%%%%%%%%%%%%%%%%%%%%%%%%%%%%%%%%

The rest parameters are each given an estimated starting value and then best-fitted by a gradient search method. First we cannot do the same for the intrinsic sodium conductance $\Nabarconductance$ as we did for the intrinsic potassium conductance $\Kbarconductance$ because we know now that Hodgkin and Huxley's purportedly sodium $IV$-curve is an aggregate of the sodium and the gating channels. The estimate for the potassium conductance is reliable but not for the sodium conductance. Because of our estimation above on the order of magnitude for the intrinsic conductances of the potassium and sodium channels we will choose a starting value for the latter to be $\Nabarconductance=100$.

As for the gating parameter, we start the initial guess for the gating resting potential at $\Gabattery=-52.5$mV, 5mV below the estimated membrane resting potential $E_r=-47.5$mV. As for the intrinsic gating conductance we will start it at $\Gbarconductance=10$, a value between $\Kbarconductance$ and $\Nabarconductance$.

As for the activation and gating range parameters we will start them at $\Nabconstant=10,\ \Kbconstant=10,\ \Gbconstant=8$. This translates to the reversal of activation for the sodium channel at $\Naabattery-\Nabconstant=49.5$mV, a non-accessible and thus unimportant potassium criticality $\Kabattery-\Kbconstant=-69.5$mV, and a gating `threshold' criticality $\Gabattery+\Gbconstant=-44.5$mV.

Last, for the adaptation time constants we use $\KTimeConstant=1$/msec, and $\NaTimeConstant=10$/msec, making the sodium conductance adaptation one order faster than the potassium conductance.

The HHAxon17 data was obtained by Hodgkin and Huxley by an instantaneous depolarization above the resting potential of the axon that translates to the initial voltage value $V(0)=-20.6707$mV. The initial values for the rest of variables are $n(0)=e^{(E_r-\Kabattery)/\Kbconstant}$, $m(0)=e^{(E_r-\Naabattery)/\Nabconstant}$, $h(0)=e^{-(E_r-\Gabattery)/\Gbconstant}$, which are fixed throughout the best fit process because they are good approximations for all searches at the initial time $t=0$.

A best fit of the model to the data is done by Newton's line search method (c.f. \cite{deng2014male,dengHLCT}). the error function is defined to be
\[
\mathcal E=\frac{1}{N}\sqrt{\sum_{j=1}^N\left[\frac{|V(t_i)-V_i|}{\bar V}\right]^2}
\]
where $\bar V$ is maximum of the absolute values of the voltage data, $N$ is the number of data points included for the error function. Thus $\mathcal E$ measures the per-data-point relative error between the predicted value $V(t_i)$ by the model and the observed value $V_i$ at time $t_i$. Fitted points include only these: the end points, the maximal and the minimal points, and the inflection points, constituting the so-called second-order fit. (The first order fit by definition would exclude the inflection points of the data.) The line search is to find the so-called best-fitted parameter values from the starting parameter values that gives rise to a local minimum point of the error function $\mathcal E$. At any iteration of the search, the error is calculated at a discrete set of points from an interval of each parameter and the new starting parameter point is chosen if it defines the smallest error. The interval is centered at the parameter value with the radius of the absolute value of the parameter, i.e. either $[0,2p]$ or $[2p,0]$ depending on if the current parameter value $p$ is positive or negative. Fig.\ref{figBestModelFit} shows both a best fit of the model and the original and a best fit of the HH model to the same data. This preliminary comparison suggests our model does better than the HH model.

\bigskip
\noindent{\bf 3. Discussion.} Mathematical modeling is a process of falsification and refinement. The HH model has become a benchmark for the squid giant axon in particular and a template for excitable membranes in general. To replace its benchmark status by our circuit model, detailed comparisons between the two are needed and they are given below.

\begin{figure}[t!]

\centerline{
\parbox[l]{2.5in}{
\centerline
{\scalebox{.5}{\includegraphics{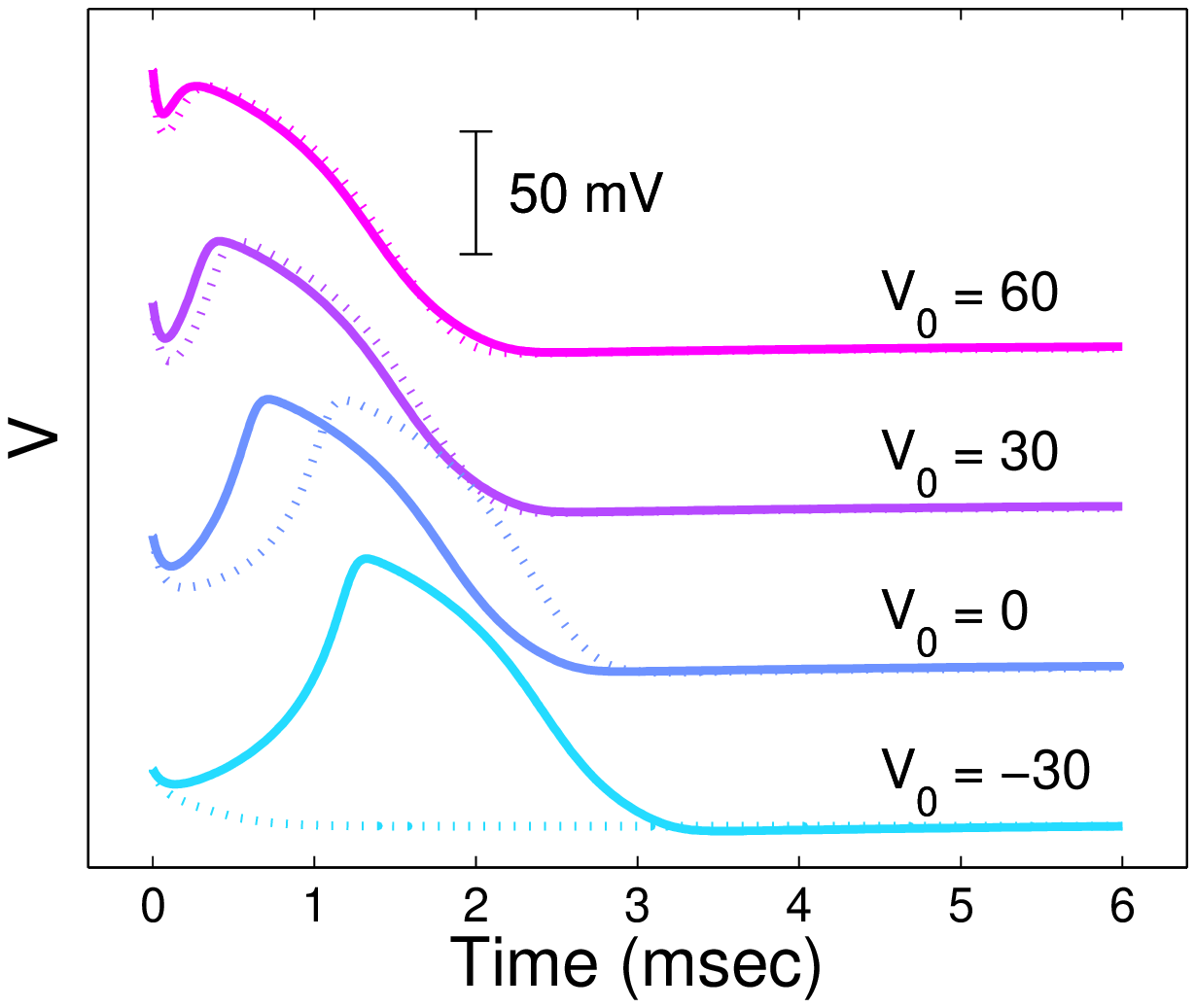}}}
}
\hskip 2cm
\parbox[l]{2.5in}{
\centerline
{\scalebox{.5}{\includegraphics{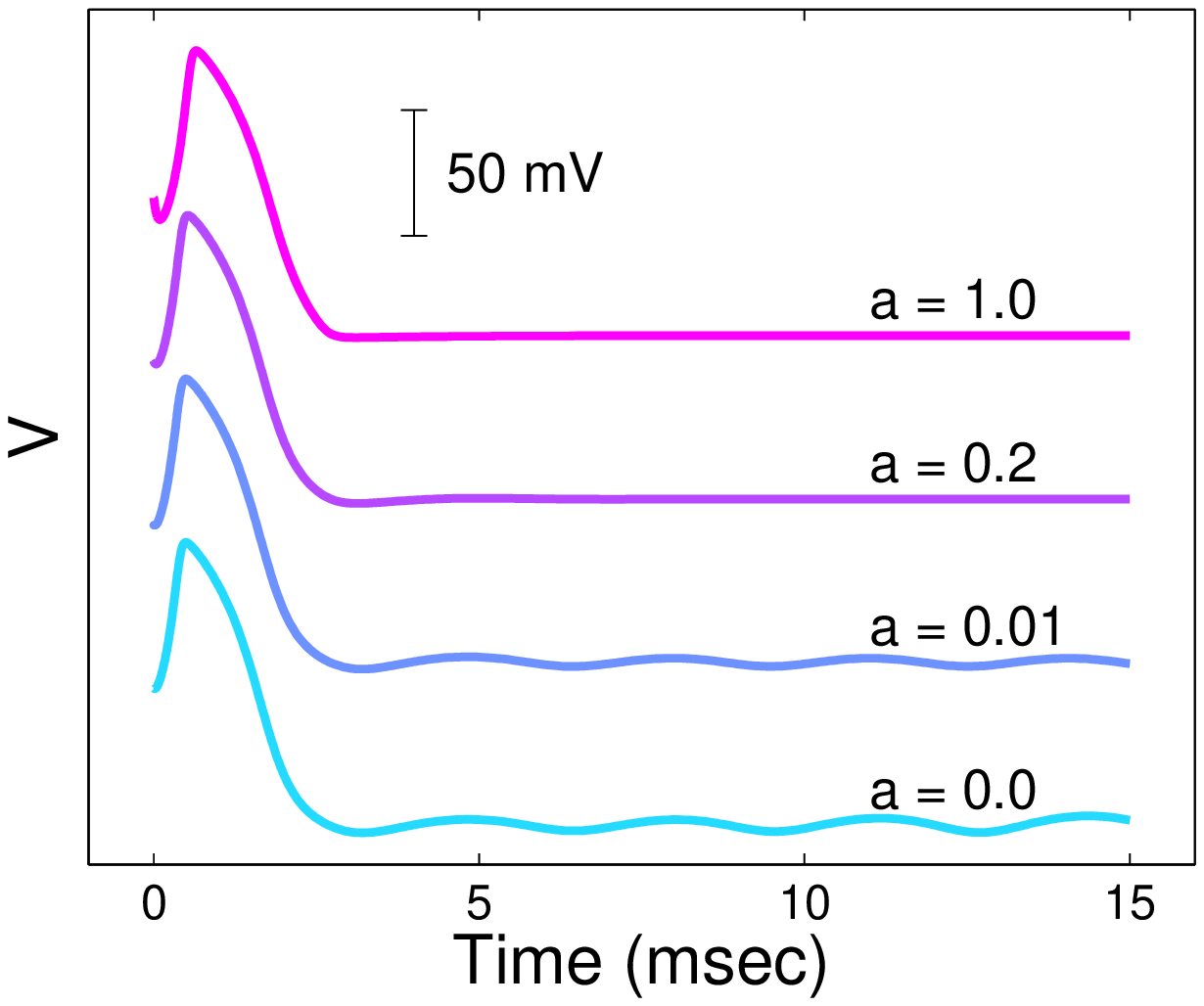}}}
}
}
\vskip .07in
\centerline{\ \ \ \ (a)\hskip 3.16in (b)}

\caption{(a) The phenomenon of the so-called absolute refraction by \cite{hodgkin1952quantitative} for different depolarizing initial voltage with the same parameter values from Fig.\ref{figBestModelFit} for model (\ref{eqDengModel}). Dotted curves are the same plot except for a larger intrinsic conductance $\Gbarconductance=15$ for gating. (b) Anode break excitation oscillations for small $0<a<1$ for equations (\ref{eqDengModel1}) with the same parameter values as (a) with $\Gbarconductance=15$ except for the new additions $a$ and  $\GTimeConstant=0.5$.}\label{figAbsoluteRefractionAnodeBreak}
\end{figure}

\medskip\noindent{\em Absolute Refraction and Anode Break Excitation.} The HH model was hugely successful. Its success was supported among other things by its matching up two uncanny properties of Hodgkin and Huxley's experiment data, the phenomenon of absolute refraction and the anode break excitation oscillation. The former occurs when a sudden initial depolarizing voltage is applied the membrane voltage decreases first before increasing or depolarizing as shown by both the axon 17 data and the models from Fig.\ref{figBestModelFit}. This phenomenon was later identified to be the phenomenon of voltage-gating (\cite{armstrong1971interaction}). It is validated by our model (\ref{eqDengModel}) as demonstrated in Fig.\ref{figAbsoluteRefractionAnodeBreak}(a). It shows not only model (\ref{eqDengModel}) exhibits the same property but also its affect by the intrinsic gating conductance $\Gbarconductance$ as it is supposed to be. In contrast, the HH model exhibits the same but does not explain it.

%%%%%%%%%%%%%%%%%%%%%%%%%%%%%%%%%%%%%%%%%%%%%%%%%%%%%%%%%%%%%
%%%%%%%%%%%%%%%%%%%%%%%%%%%%%%%%%%%%%%%%%%%%%%%%%%%%%%%%%%%%%
%%%%%%%%%%%%%%%%%%%%%%%%%%%%%%%%%%%%%%%%%%%%%%%%%%%%%%%%%%%%%
%%%%%%%%%%%%%%%%%%%%%%%%%%%%%%%%%%%%%%%%%%%%%%%%%%%%%%%%%%%%%
%%%%%%%%%%%%%%%%%%%%%%%%%%%%%%%%%%%%%%%%%%%%%%%%%%%%%%%%%%%%%
%%%%%%%%%%%%%%%%%%%%%%%%%%%%%%%%%%%%%%%%%%%%%%%%%%%%%%%%%%%%%

As for the anode break excitation, it refers to the phenomenon that after hyperpolarization and during the phase of re-polarization, the membrane voltage oscillates towards its resting potential rather than does so monotonically. The HH model is capable of this phenomenon but does not match up well in the time scale (Fig.22 of \cite{hodgkin1952quantitative}). To capture this property with the correct time scale, we make a small modification to model (\ref{eqDengModel}). We assume that a proportion of the gating current $\Gpcurrent=\Gbarconductance e^{-(V-\Gabattery)/\Gbconstant}(V-\Gabattery)$ is subject to conductance adaptation and the remaining proportion is subject to current adaptation modeled by the last equation of the system below
\begin{equation}\label{eqDengModel1}
\left\{\!\!\!\!\begin{array}{ll} & C\Vvoltageprime=-[\Kbarconductance n (\Vvoltage-\Kabattery)+\Nabarconductance m (\Vvoltage-\Naabattery)+a\Gbarconductance h (\Vvoltage-\Gabattery)+(1-a)\Gpcurrent-\Ecurrent]\\
& {n}' = \KTimeConstant(e^{(\Vvoltage-\Kabattery)/\Kbconstant}-n)\\
& {m}' = \NaGTimeConstant(e^{(\Vvoltage-\Naabattery)/\Nabconstant}-m)\\
& {h}' = \NaGTimeConstant(e^{-(\Vvoltage-\Gabattery)/\Gbconstant}-h)\\
& {\Gpcurrent}' = \GTimeConstant(\Gbarconductance e^{-(\Vvoltage-\Gabattery)/\Gbconstant} (\Vvoltage-\Gabattery)-\Gpcurrent)\\
\end{array}\right.
\end{equation}
where $a:(1-a)$ is the proportionality split for the two types of adaptation. Fig.\ref{figAbsoluteRefractionAnodeBreak} (a) shows that if the proportion for current adaptation is substantial (small $a$) but slow (small $\GTimeConstant$), anode break excitation occurs with the same time scale as shown by the experimental data of \cite{hodgkin1952quantitative} (Fig.22).
We should note that the phenomena of absolute refraction and anode break excitation are notorious for other models to replicate, the FitzHugh-Nagumo model (\cite{fitzhugh1961impulses}) or the Hindmarsh-Rose model (\cite{hindmarsh1984model}) are two such examples.

%%%%%%%%%%%%%%%%%%%%%%%%%%%%%%%%%%%%%%%%%%%%%%%%%%%%%%%%%%%%%
%%%%%%%%%%%%%%%%%%%%%%%%%%%%%%%%%%%%%%%%%%%%%%%%%%%%%%%%%%%%%
%%%%%%%%%%%%%%%%%%%%%%%%%%%%%%%%%%%%%%%%%%%%%%%%%%%%%%%%%%%%%
\begin{figure}[t!]

\centerline{
\parbox[l]{3in}{
\centerline
{\scalebox{.5}{\includegraphics{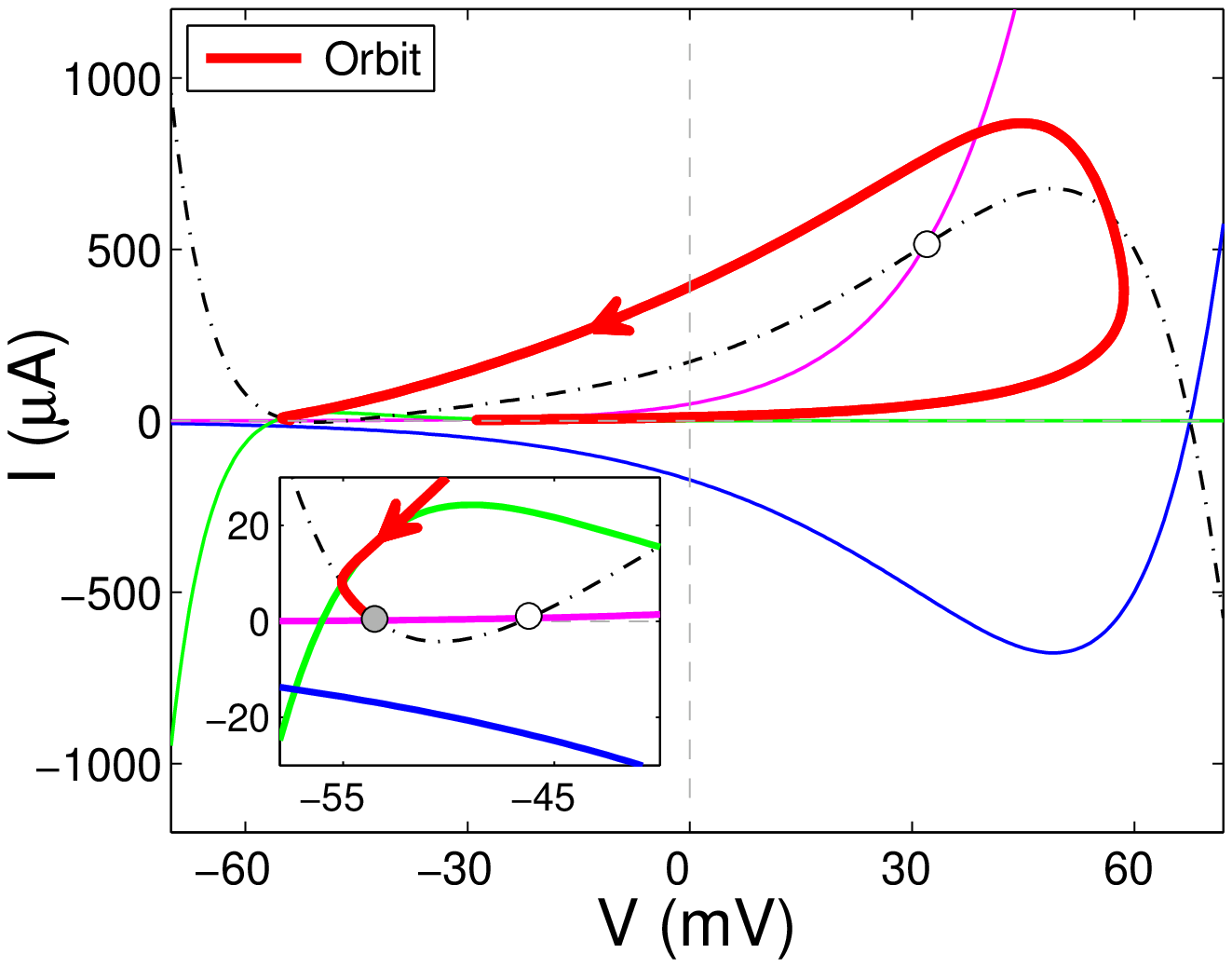}}}
}
\hskip 1cm
\parbox[l]{3in}{
\centerline
{\scalebox{.5}{\includegraphics{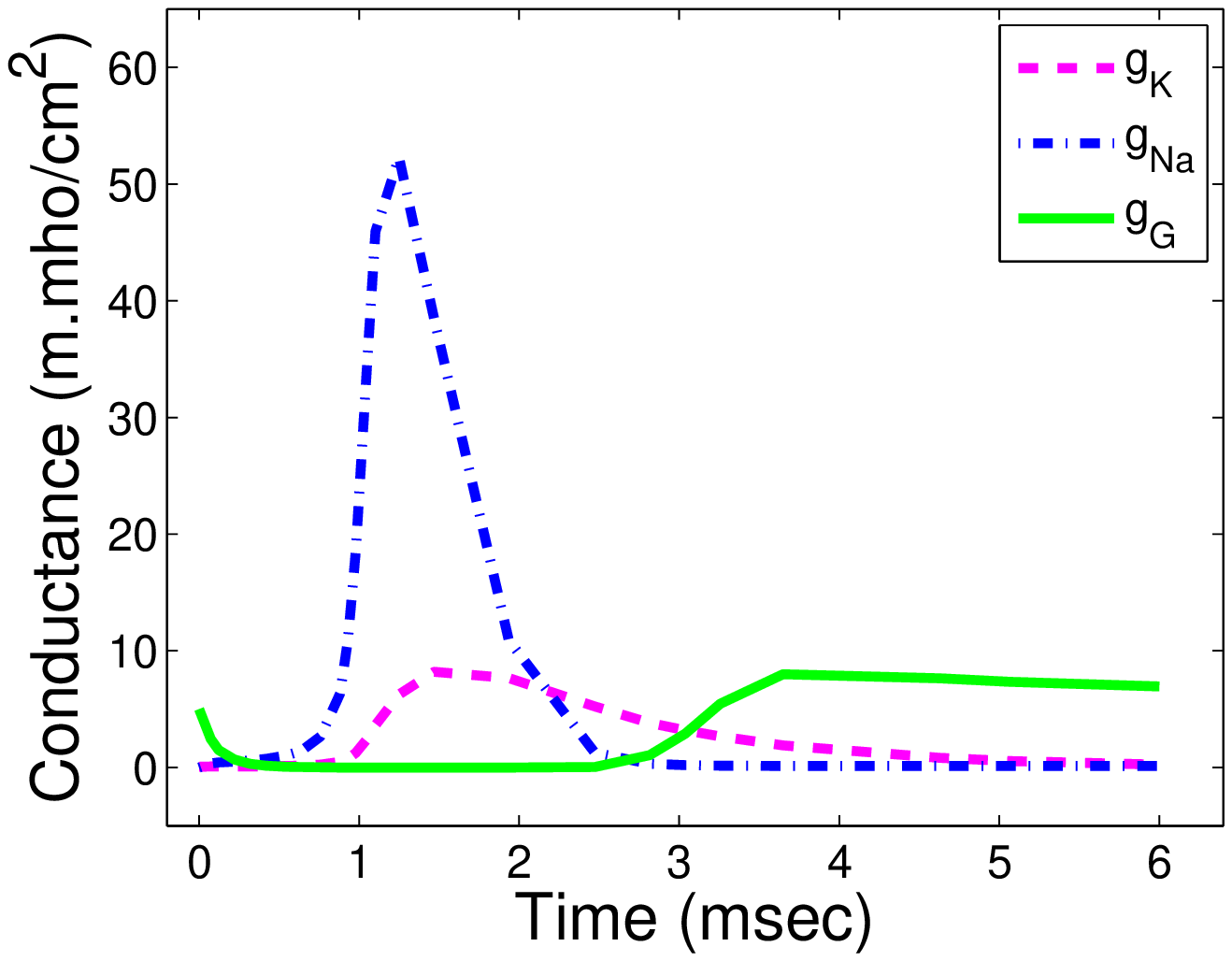}}}
}
}
\vskip .07in
\centerline{\ \ \ \ (a)\hskip 3.3in (b)}

%\vskip .1in

\centerline{
\parbox[l]{3in}{
\centerline
{\scalebox{.5}{\includegraphics{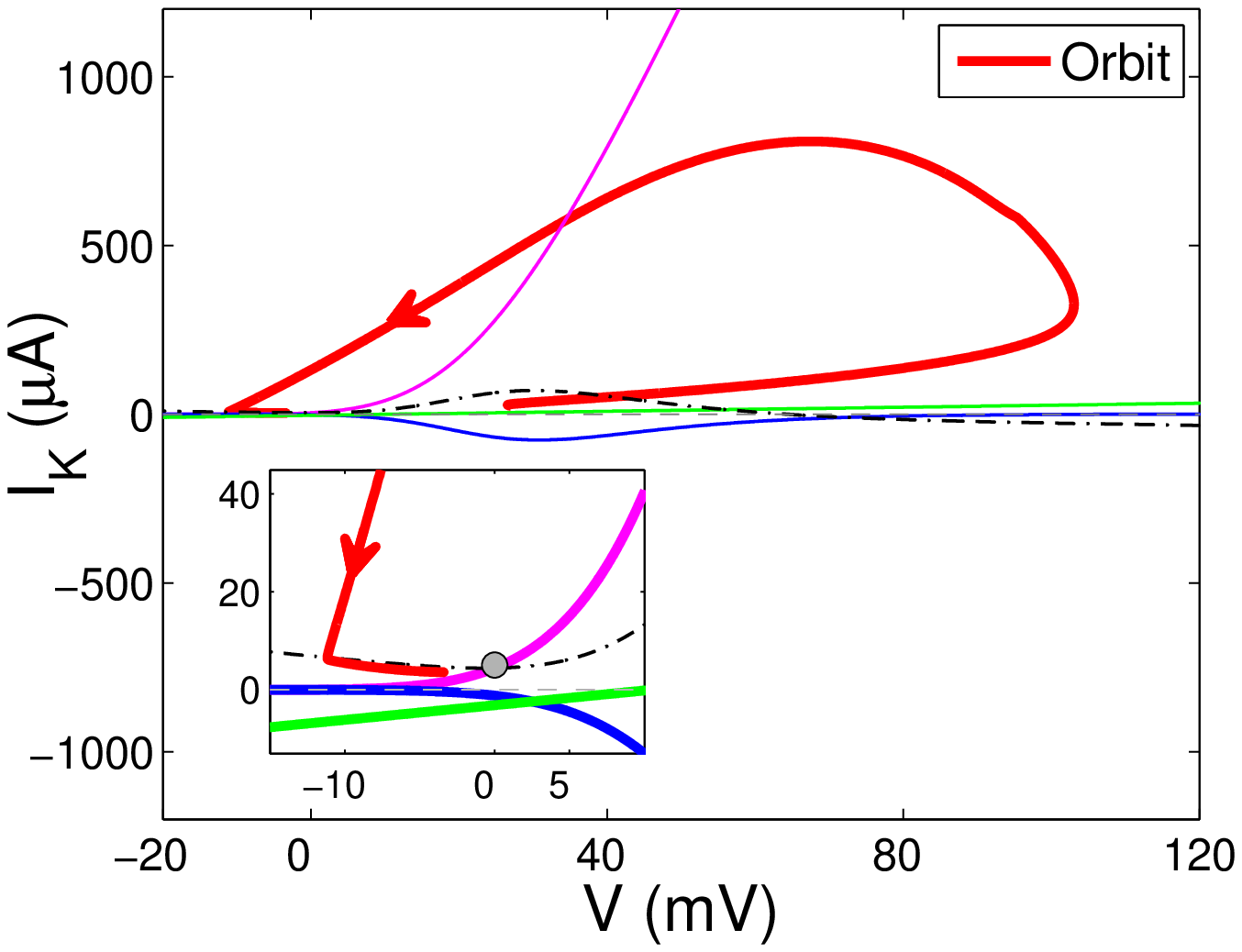}}}
}
\hskip 1cm
\parbox[l]{3in}{
\centerline
{\scalebox{.5}{\includegraphics{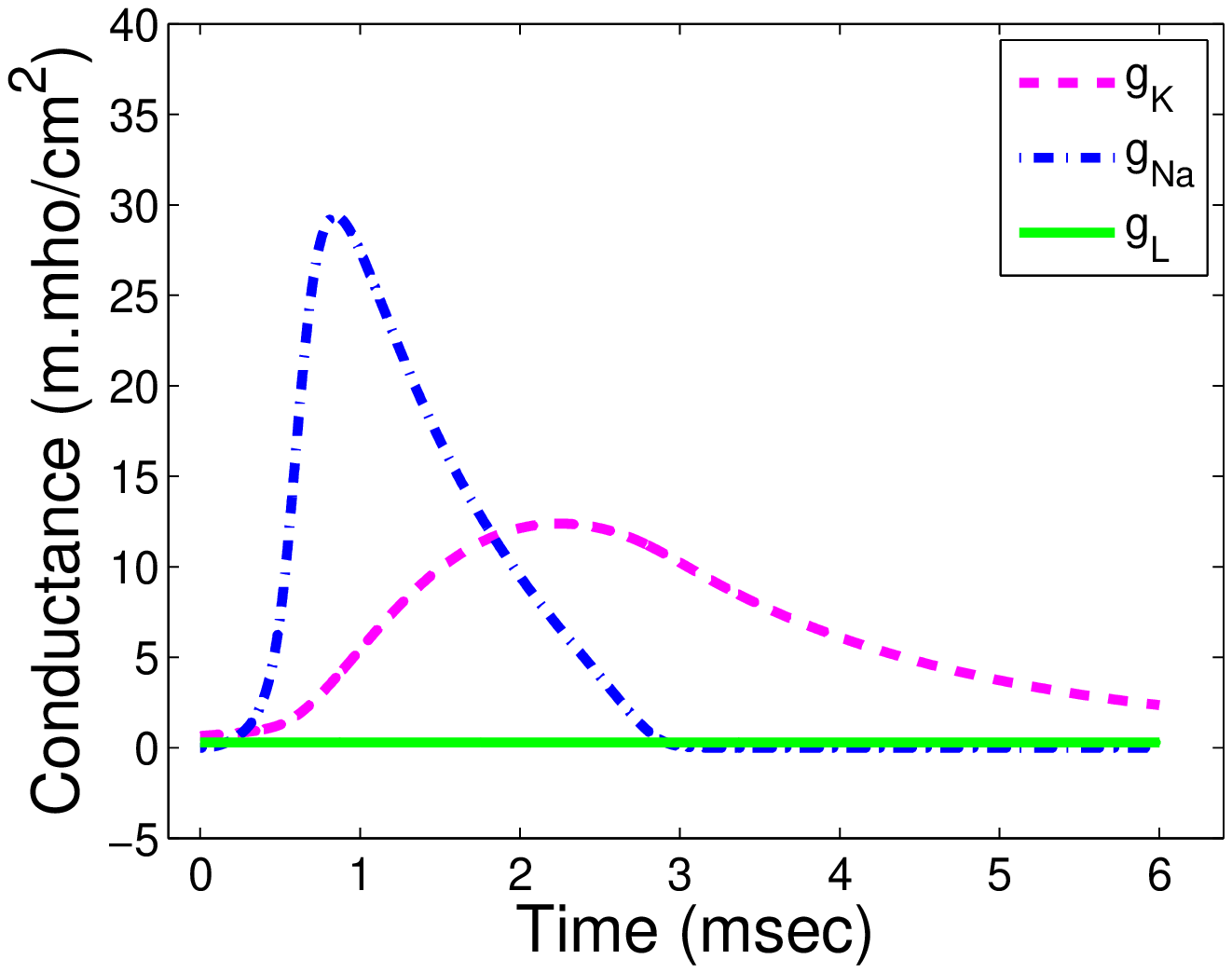}}}
}
}
\vskip .07in
\centerline{\ \ \ \ (c)\hskip 3.3in (d)}

\caption{(a) The best-fitted action potential of (\ref{eqDengModel}) is projected to the $V\Kpcurrent$-space together with its various $IV$-curves. (b) Its conductances as functions of the time for the action potential. (c) and (d) Similar plot as (a) and (b) for the HH model. The dash-dot curve is the reflection of the sodium-leakage $IV$-curve $I=-(\NaIVf(V)+\LIVf(V))$ in (c). }\label{figCompareToHHModel}
\end{figure}
%%%%%%%%%%%%%%%%%%%%%%%%%%%%%%%%%%%%%%%%%%%%%%%%%%%%%%%%%%%%%
%%%%%%%%%%%%%%%%%%%%%%%%%%%%%%%%%%%%%%%%%%%%%%%%%%%%%%%%%%%%%
%%%%%%%%%%%%%%%%%%%%%%%%%%%%%%%%%%%%%%%%%%%%%%%%%%%%%%%%%%%%%

\medskip\noindent{\em All-or-Nothing Action Potentials.} Although both the HH model and our model (\ref{eqDengModel}) match up Hodgkin and Huxley's experimental data in many aspects, many fundamental differences remain. Fig.\ref{figCompareToHHModel} shows some. The first concerns the all-or-nothing generation of action potentials. As it is shown in Fig.\ref{figCompareToHHModel}(a), in addition to the membrane resting potential $\mbattery$, our model has two more equilibrium points: one to the right of $\mbattery$, which is a saddle point, and one to its far right, which is typically a unstable spiral. It is known from the theory of dynamical systems that the resting membrane potential is always stable and the triple-equilibria configuration creates a saddle-node bifurcation point for the threshold of firing, giving rise to an all-or-nothing firing mechanism for action potentials. If the initial voltage is to the right of the middle saddle-node, an action potential ensues, with a magnitude stretching passing the spiral focus at the least. In contrast, Fig.\ref{figCompareToHHModel}(c) shows that the HH model has only one equilibrium point which is the resting membrane potential $\mbattery$. The corresponding firing mechanism is by the way of a Hopf-bifurcation point. In theory, the action potentials are graded, not the all-or-nothing type as is supposed to be. That is, depending on how close the initial voltage is to the Hopf-point (the local minimum point of the reflection of the joint sodium-leakage $IV$-curve $I=-(\NaIVf(V)+\LIVf(V))$), the magnitude of an action potential can vary from nothing to full, a phenomenon of the so-called canard explosion (\cite{benoit1981chasse,deng2004food}).

%%%%%%%%%%%%%%%%%%%%%%%%%%%%%%%%%%%%%%%%%%%%%%%%%%%%%%%%%%%%%
%%%%%%%%%%%%%%%%%%%%%%%%%%%%%%%%%%%%%%%%%%%%%%%%%%%%%%%%%%%%%
%%%%%%%%%%%%%%%%%%%%%%%%%%%%%%%%%%%%%%%%%%%%%%%%%%%%%%%%%%%%%
\begin{figure}[t!]

\centerline{
\parbox[l]{2.5in}{
\centerline
{\scalebox{.5}{\includegraphics{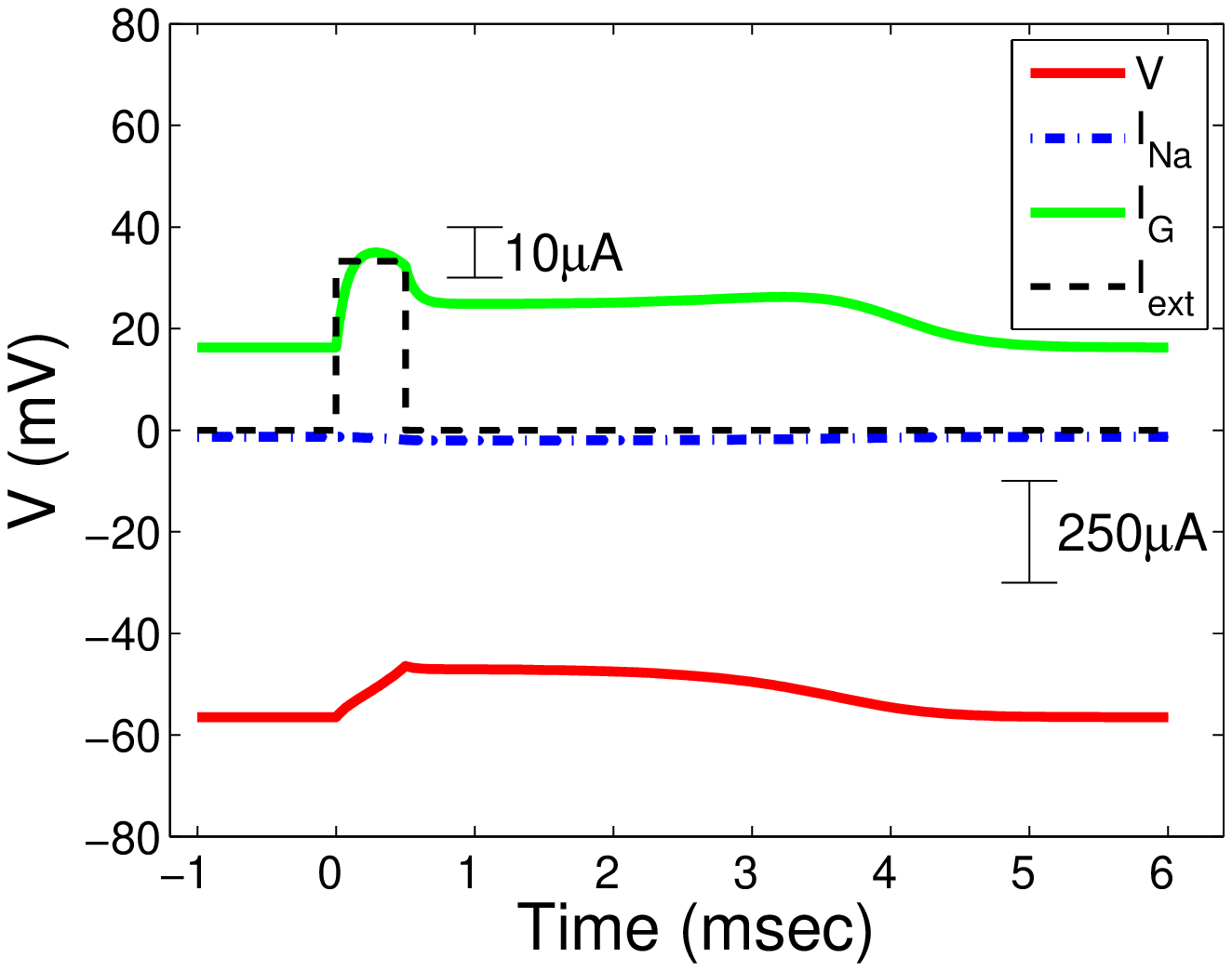}}}
}
\hskip 2cm
\parbox[l]{2.5in}{
\centerline
{\scalebox{.5}{\includegraphics{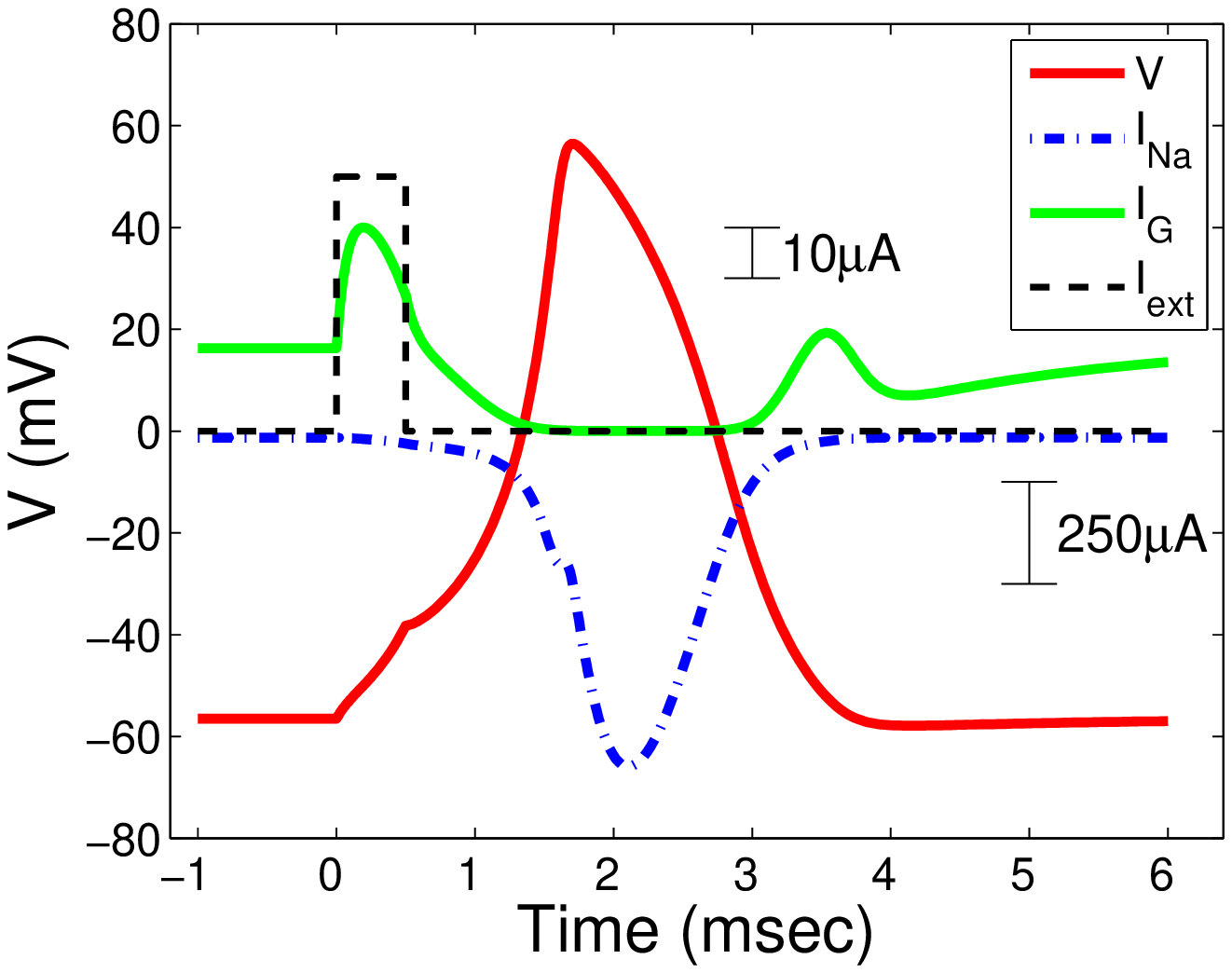}}}
}
}
\vskip .07in
\centerline{\ \ \ \ (a)\hskip 3.16in (b)}

\centerline{
\parbox[l]{2.5in}{
\centerline
{\scalebox{.5}{\includegraphics{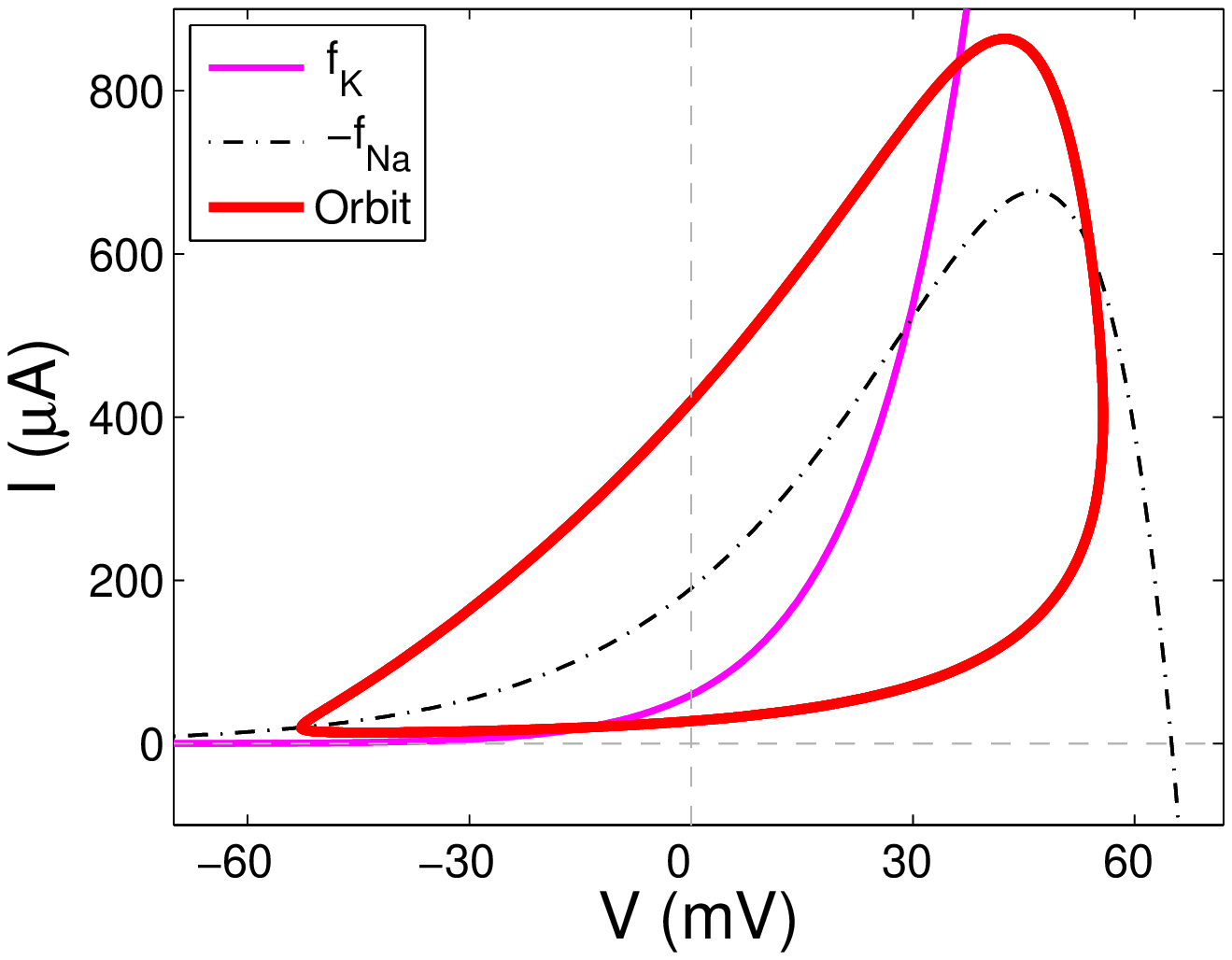}}}
}
\hskip 2cm
\parbox[l]{2.5in}{
\centerline
{\scalebox{.5}{\includegraphics{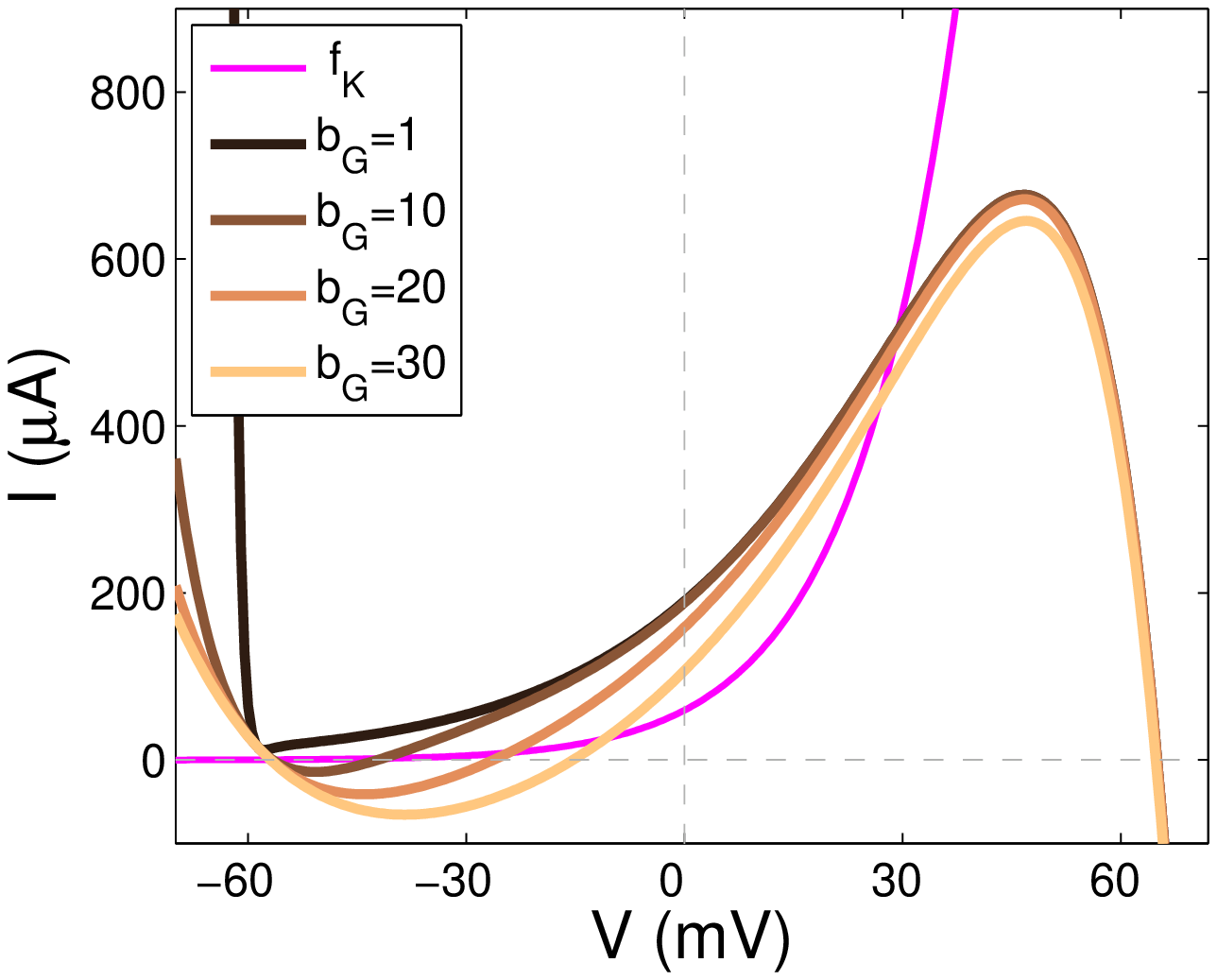}}}
}
}
\vskip .07in
\centerline{\ \ \ \ (c)\hskip 3.16in (d)}

\caption{(a) With the same parameter values as from Fig.\ref{figBestModelFit}(a) for model (\ref{eqDengModel}) but forced by an excitatory square pulse $\Ecurrent$ just below the action potential threshold which is about the maximal value of the gating current. (b) The same plot except for an above-threshold square pulse. (c) A cyclic firing without the repolarizing resting potential if voltage-gating is absent ($\Gbarconductance=0$). (d) The sodium-gating $IV$-curves with varying gating range $\Gbconstant$. Too small or too big a $\Gbconstant$ is bad for action potential generation.}\label{figGating}
\end{figure}
%%%%%%%%%%%%%%%%%%%%%%%%%%%%%%%%%%%%%%%%%%%%%%%%%%%%%%%%%%%%%
%%%%%%%%%%%%%%%%%%%%%%%%%%%%%%%%%%%%%%%%%%%%%%%%%%%%%%%%%%%%%
%%%%%%%%%%%%%%%%%%%%%%%%%%%%%%%%%%%%%%%%%%%%%%%%%%%%%%%%%%%%%

There are three more noticeable differences about the firing mechanism. For our model, the stability of the resting potential is due to the voltage-gating $IV$-curve, whereas that for the HH model is  critically dependent of the presence of the leakage current $\Lpcurrent$. Second, the re-polarization of the membrane for our model is due to the voltage-gating characteristics as the sodium channel is closing down when charged molecules return to the sodium gate pores. In contrast, the re-polarization of the membrane by the HH model is carried out by the leakage channel. Third, our model predicts a persisting leaking current when the membrane is at the resting potential $\mbattery$ as shown in the inset of  Fig.\ref{figCompareToHHModel}(a) of which the primary components are the gating and the sodium currents, cancelling out each other at the resting equilibrium. That is, the leakage is a consequence of the voltage-gating rather than a cause by the HH model shown in the inset of  Fig.\ref{figCompareToHHModel}(c).

%%%%%%%%%%%%%%%%%%%%%%%%%%%%%%%%%%%%%%%%%%%%%%%%%%%%%%%%%%%%%
%%%%%%%%%%%%%%%%%%%%%%%%%%%%%%%%%%%%%%%%%%%%%%%%%%%%%%%%%%%%%
%%%%%%%%%%%%%%%%%%%%%%%%%%%%%%%%%%%%%%%%%%%%%%%%%%%%%%%%%%%%%
\begin{figure}[t!]

\centerline{
\parbox[l]{2.5in}{
\centerline
{\scalebox{.5}{\includegraphics{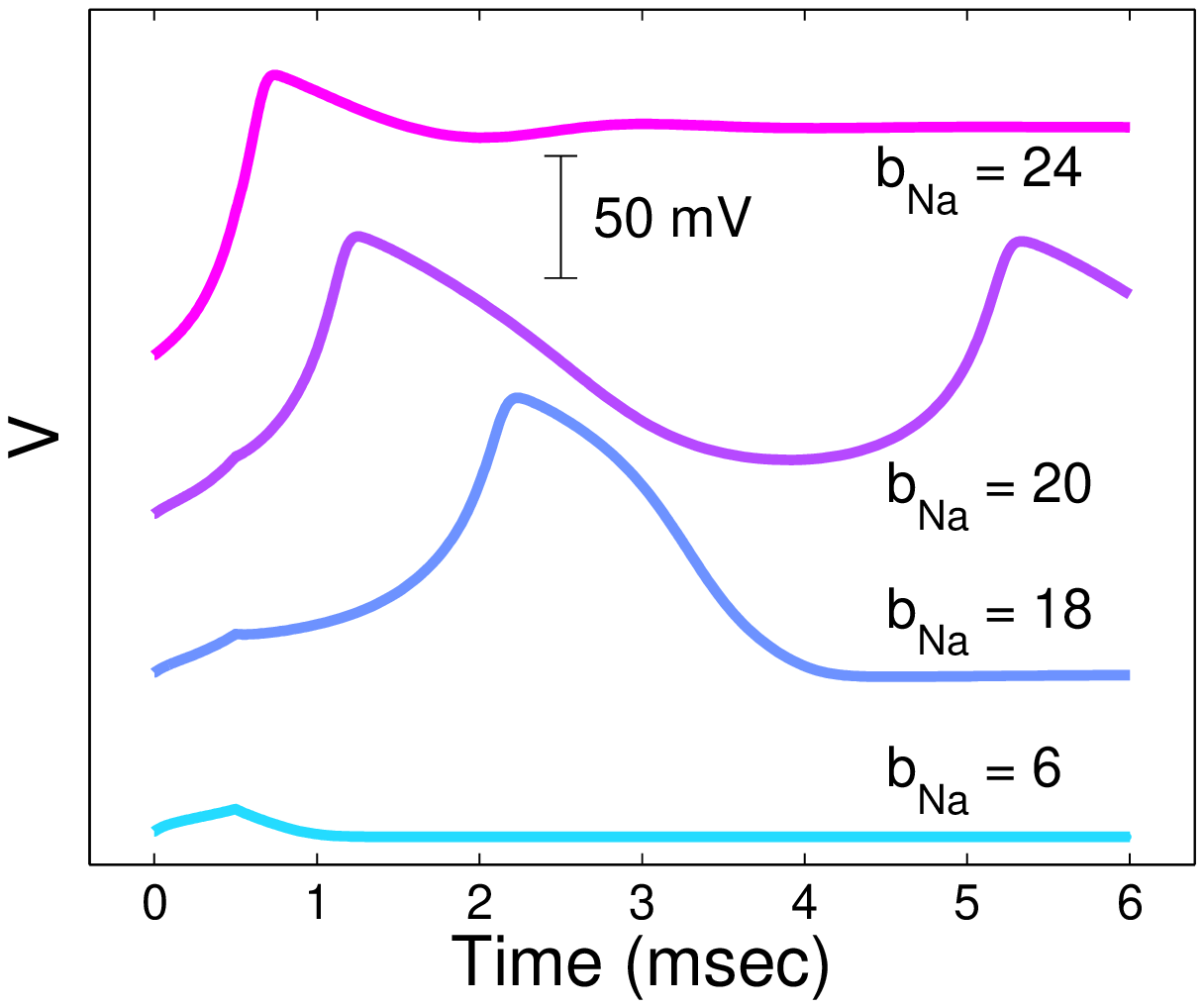}}}
}
\hskip 2cm
\parbox[l]{2.5in}{
\centerline
{\scalebox{.5}{\includegraphics{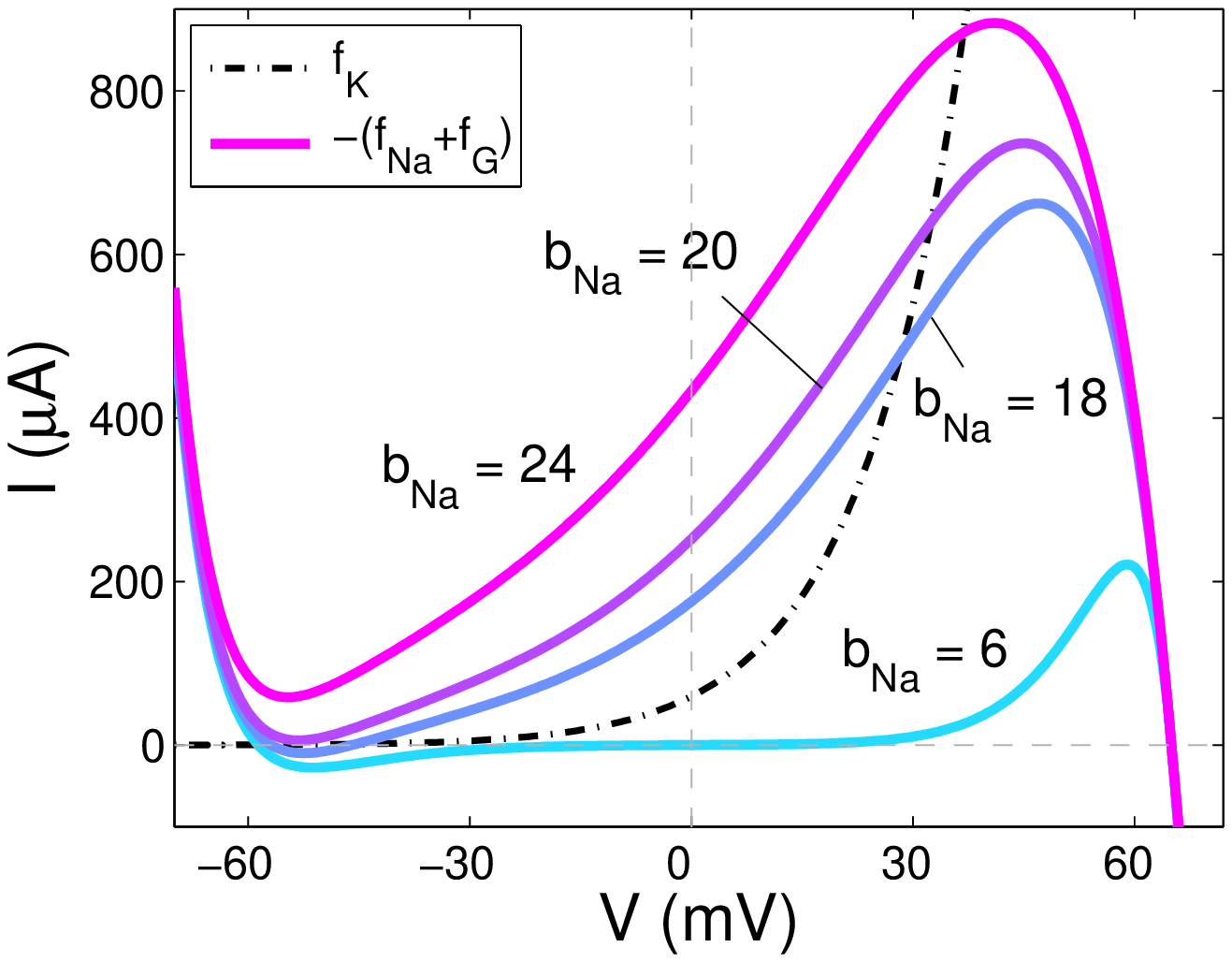}}}
}
}
\vskip .07in
\centerline{\ \ \ \ (a)\hskip 3.16in (b)}

\caption{(a) With the same initial conditions, the same parameter values, and the same excitation forcing as in  Fig.\ref{figGating}(b) but with varying sodium activation range parameter $\Nabconstant$. (b) The corresponding $IV$-curve configurations. Too small or too big a $\Nabconstant$ is not desirable for action potentials.}\label{figActivation}
\end{figure}
%%%%%%%%%%%%%%%%%%%%%%%%%%%%%%%%%%%%%%%%%%%%%%%%%%%%%%%%%%%%%
%%%%%%%%%%%%%%%%%%%%%%%%%%%%%%%%%%%%%%%%%%%%%%%%%%%%%%%%%%%%%
%%%%%%%%%%%%%%%%%%%%%%%%%%%%%%%%%%%%%%%%%%%%%%%%%%%%%%%%%%%%%

As for the action potential at large, our model shows it tracks closely around the sodium-gating curve $I=-(\NaIVf(V)+\GIVf(V))$. Also the reversal of the potassium current (trajectory entering the left side of the potassium $IV$-curve) happens rather promptly after the potassium current has peaked. In contrast, the action potential of the HH model comes close only to two points of the $IV$-curves, the Nernst potentials $\Kabattery,\ \Naabattery$, and with the rest playing little role to guide the trajectory. And its potassium current reversal is rather delayed after the potassium current has peaked and the membrane has started depolarization.

\medskip\noindent{\em Voltage-Gating.} One major difference lies in the inclusion of voltage-gating into our model. As shown in  Fig.\ref{figCompareToHHModel}(b), voltage-gating is active during the initialization phase and the termination phase of an action potential. Both are for the purpose of preventing (accidental) firing of action potentials, and yet once the firing threshold is exceeded, the gating conductance recedes, permitting the membrane to depolarize. In contrast, little can be deduced from the leakage conductance time course (which is a constant of time) from the HH model as shown in Fig.\ref{figCompareToHHModel}(d). Fig.\ref{figGating} also shows more clearly what happens with the injection of an excitatory intracellular square pulse $\Ecurrent(t)$. Voltage-gating suppresses an action potential if the excitation fails to clear the threshold, approximately the maximum of the gating current. Otherwise an action potential is generated if the excitation clears the threshold. The figure also shows that taking away gating ($\Gbarconductance=0$), the membrane resting potential $\mbattery$ is gone for losing its stability and the action potential is replaced by a perpetual limit cycle. It also shows that if the gating range is too close to the gating resting potential $\Gabattery$ (very small $\Gbconstant$) it is effectively the same as if there is no gating ($\Gbarconductance=0$). Also if the range of gating is far away (large $\Gbconstant$) the threshold for action potential can become too high for intracellular excitation to clear. There is a `Goldilocks Zone' from the resting gating potential for the range of gating.

\medskip\noindent{\em Ion Current Activation.} Fig.\ref{figActivation} shows the effect of the sodium current activation range parameter $\Nabconstant$ on the generation of action potentials. If the sodium activation is too close to its Nersnt resting potential (small $\Nabconstant$) it requires a very long excitation time for action potentials to arise. If the activation is far away (large $\Nabconstant$), it renders the voltage-gating useless. It also requires a `Goldilocks Zone' from the resting sodium potential for the range of sodium activation.

\medskip\noindent{\em Model Overfit.} Another major difference lies in the arbitrariness of the functional forms for the voltage-dependent conductances of the HH model. For example, for the function $\alpha_n$ of the HH model, four parameters were actually required and fitted to have its final form obtained in \cite{hodgkin1952quantitative} because it is of this general form
\[
\alpha_n=\frac{a_1-V}{a_2(\exp(a_3-a_4V)-1)}.
\]
Similarly, the function $\beta_n$ was fitted with two free parameters. Altogether, the HH model had 13 more parameters for its fitting than our model (of which only the $b$-parameters and the $\tau$-parameters are not shared with the HH model). Because the functional forms of the HH model and their parameters are rather arbitrary, its fit to the data can be construed as `overfit' as with the case where arbitrary polynomials can fit but not explain any data.

\begin{figure}[t!]

\centerline{
\parbox[l]{2.5in}{
\centerline
{\scalebox{.5}{\includegraphics{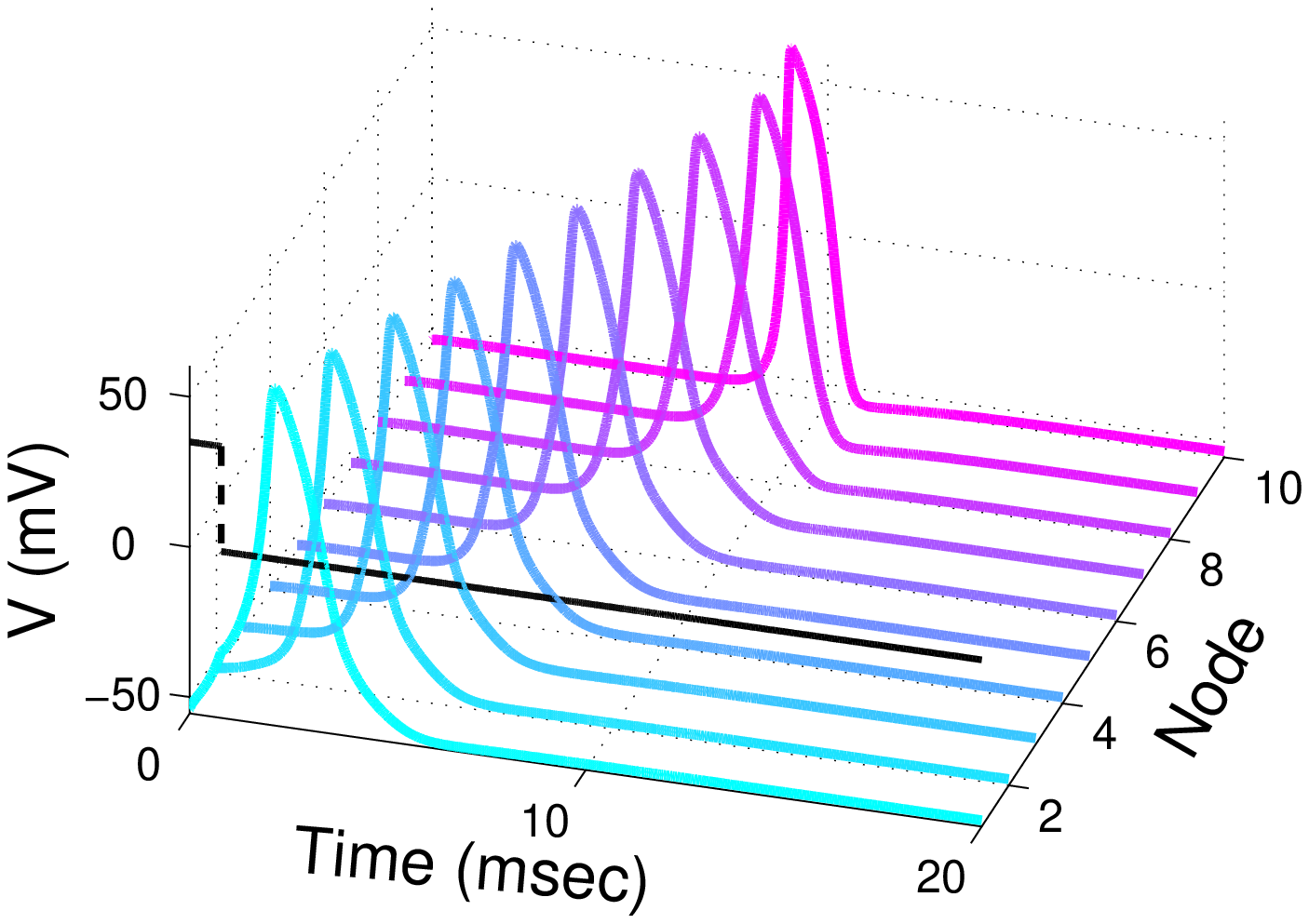}}}
}
\hskip 2cm
\parbox[l]{2.5in}{
\centerline
{\scalebox{.5}{\includegraphics{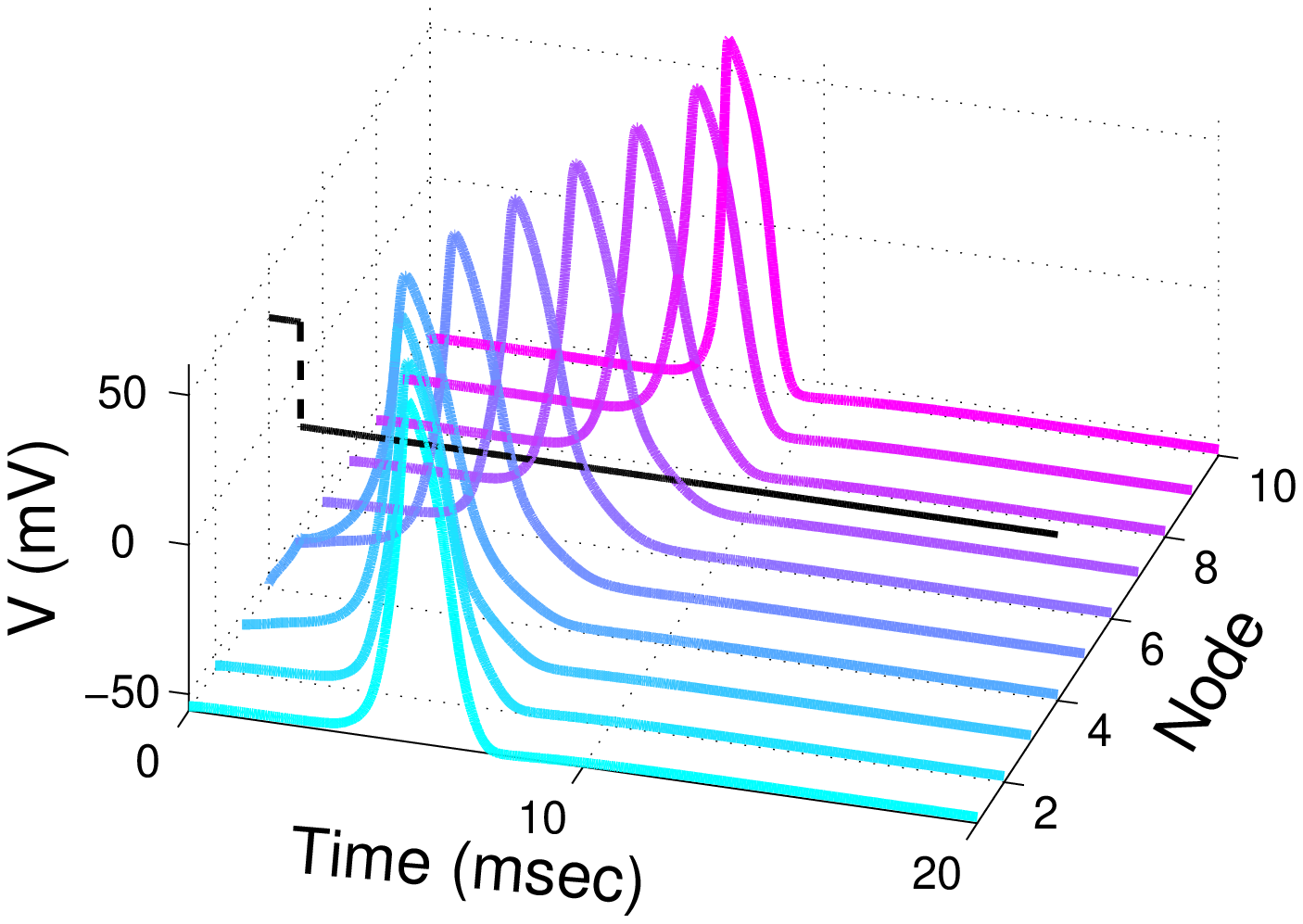}}}
}
}
\vskip .07in
\centerline{\ \ \ \ (a)\hskip 3.16in (b)}

\caption{With the same parameter values as from Fig.\ref{figBestModelFit}(a) for model (\ref{eqDengModelMyelinated}), and $d=0.6$ m.mho per square centimeter, (a) shows the traveling pulse if the first node is injected by an intracellular current $\Ecurrent=35$ $\mu$A for the first $0.8$ msec. Plot (b) shows the same except that the excitatory current is injected at the fourth node.}\label{figTravelingWave}
\end{figure}

\medskip\noindent{\em Traveling Action Potential.} The term action potential in its original definition is referred to the uniform profile of the membrane voltage when an electrical pulse propagates from one end of the axon to another. The partial differential equations for such propagating pulse is derived by adding the axial diffusive current $DV_{xx}$ to the total current at each location to the patch model (\ref{eqDengModel}). The derivation follows the same treatment as in \cite{hodgkin1952quantitative}. The spatially continuous model is
\begin{equation}\label{eqDengModelAxon}
\left\{\!\!\!\!\begin{array}{ll} & CV_t=DV_{xx}-[\Kbarconductance n (\Vvoltage-\Kabattery)+\Nabarconductance m (\Vvoltage-\Naabattery)+\Gbarconductance h (\Vvoltage-\Gabattery)-\Ecurrent]\\
& {n}_t = \KTimeConstant(e^{(\Vvoltage-\Kabattery)/\Kbconstant}-n)\\
& {m}_t = \NaGTimeConstant(e^{(\Vvoltage-\Naabattery)/\Nabconstant}-m)\\
& {h}_t = \NaGTimeConstant(e^{-(\Vvoltage-\Gabattery)/\Gbconstant}-h)\\
\end{array}\right.
\end{equation}
with $D$ being the axial diffusion coefficient of the axon. One can also derive a discrete version of the continuum model as follows
\begin{equation}\label{eqDengModelMyelinated}
\left\{\!\!\!\!\begin{array}{ll} & CV_i'=-[\Kbarconductance n_i (V_i-\Kabattery)+\Nabarconductance m_i (V_i-\Naabattery)+\Gbarconductance h_i (V_i-\Gabattery)\\
&\qquad\qquad\quad -\Ecurrent+d_i(V_i-V_{i-1})-d_{i+1}(V_{i+1}-V_i)]\\
& {n}_i' = \KTimeConstant(e^{(V_i-\Kabattery)/\Kbconstant}-n_i)\\
& {m}_i' = \NaGTimeConstant(e^{(V_i-\Naabattery)/\Nabconstant}-m_i)\\
& {h}_i' = \NaGTimeConstant(e^{-(V_i-\Gabattery)/\Gbconstant}-h_i)\\
\end{array}\right.
\end{equation}
with $i=1,\dots,n$, $d_1=d_{n+1}=0$, and $d_i=d$, a constant for $i=2,\dots,n$. The latter can also be used as a model for myelinated axon with $i$ denoting the nodes of Ranvier ordered from one end of the axon to another. Fig.\ref{figTravelingWave} shows the phenomenon of saltatory propagation by the discrete model.

%%%%%%%%%%%%%%%%%%%%%%%%%%%%%%%%%%%%%%%%%%%%%%%%%%%%%%%%%%%%%
%%%%%%%%%%%%%%%%%%%%%%%%%%%%%%%%%%%%%%%%%%%%%%%%%%%%%%%%%%%%%
%%%%%%%%%%%%%%%%%%%%%%%%%%%%%%%%%%%%%%%%%%%%%%%%%%%%%%%%%%%%%
%%%%%%%%%%%%%%%%%%%%%%%%%%%%%%%%%%%%%%%%%%%%%%%%%%%%%%%%%%%%%
%%%%%%%%%%%%%%%%%%%%%%%%%%%%%%%%%%%%%%%%%%%%%%%%%%%%%%%%%%%%%
%%%%%%%%%%%%%%%%%%%%%%%%%%%%%%%%%%%%%%%%%%%%%%%%%%%%%%%%%%%%%

\medskip\noindent{\em Spike Burst.} We end this section with the phenomenon of spike burst (\cite{arvanitaki1949prototypes,li1953microelectrode,frazier1967morphological,carpenter1970dependence}) for which the membrane potential of an excitable cell sustains a number of rapid spikes before hyperpolarizing and then re-polarizing near the resting potential of the membrane. Early models for bursting spikes include
\cite{connor1971prediction,partridge1976mechanism,morris1981voltage,chay1983minimal,hindmarsh1984model}.
We will not fit our model to any specific data per se but rather highlight the simple mechanisms and configurations for spike bursting. The basic template for spike burst requires the all-or-nothing firing mechanism shown in Fig.\ref{figCompareToHHModel}(a). The key difference between the one-spike action potential and the many-spikes burst is the time adaptation constant for the potassium conductance $\KTimeConstant$. For the former it has a slower time scale as shown in the figure where the sodium-gating reflection curve $I=-(\NaIVf(V)+\GIVf(V))$ strongly pulls the orbit to the left side of the potassium $IV$-curve $I=\KIVf(V)$. If we speed up the potassium adaptation by increasing its time constant, the orbit will come down faster towards the potassium $IV$-curve $I=\KIVf(V)$, and as a result oscillate around the spiral focus equilibrium point at the far right. That is, the middle saddle-node equilibrium can act to direct the orbit either to its left if the time constant $\KTimeConstant$ is small or to its right to form spike burst if $\KTimeConstant$ is large. Together, this forms the basic ingredients for spike burst.

%%%%%%%%%%%%%%%%%%%%%%%%%%%%%%%%%%%%%%%%%%%%%%%%%%%%%%%%%%%%%
%%%%%%%%%%%%%%%%%%%%%%%%%%%%%%%%%%%%%%%%%%%%%%%%%%%%%%%%%%%%%
%%%%%%%%%%%%%%%%%%%%%%%%%%%%%%%%%%%%%%%%%%%%%%%%%%%%%%%%%%%%%
\begin{figure}%[ht]

\centerline{
\parbox[l]{2.5in}{
\centerline
{\scalebox{.5}{\includegraphics{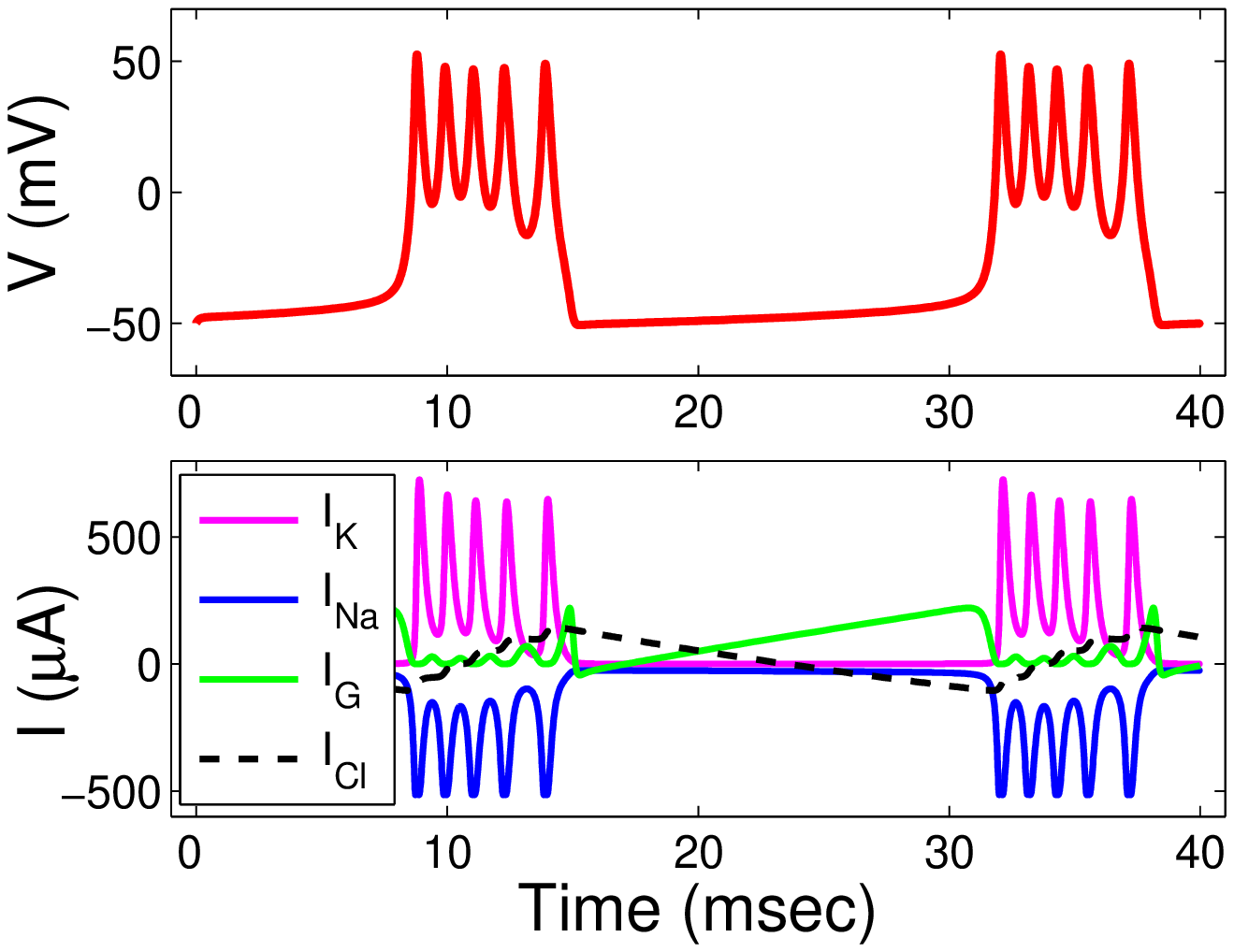}}}
}
\hskip 2cm
\parbox[l]{2.5in}{
\centerline
{\scalebox{.5}{\includegraphics{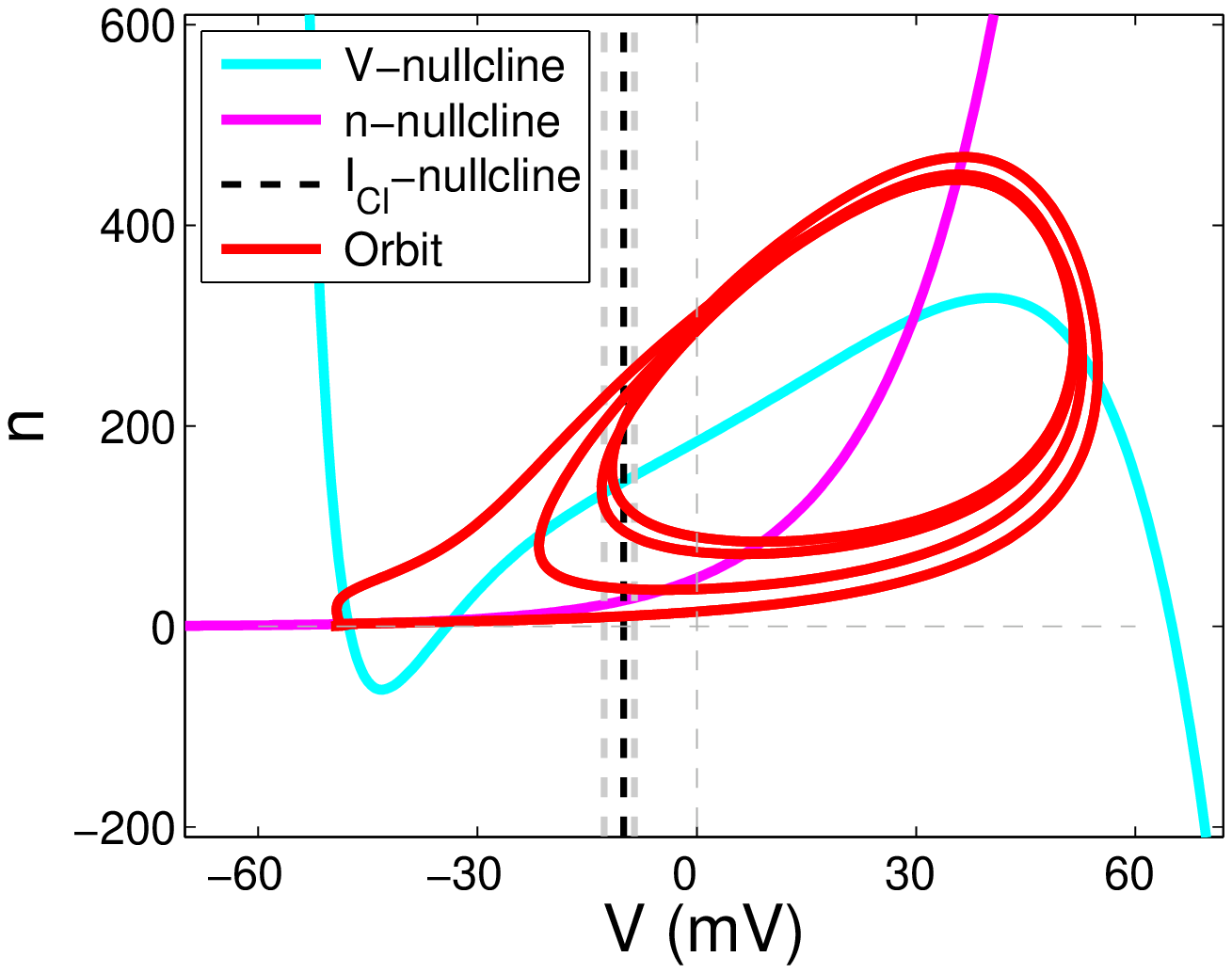}}}
}
}
\vskip .07in
\centerline{\ \ \ \ (a)\hskip 3.16in (b)}

%\centerline{
%\parbox[l]{2.5in}{
%\centerline
%{\scalebox{.5}{\includegraphics{figp2chSpikeDModelTimeSeries.eps}}}
%}
%\hskip 1cm
%\parbox[l]{2.5in}{
%\centerline
%{\scalebox{.5}{\includegraphics{figp2chSpikeDModelPhasePortrait.eps}}}
%}
%}
%\vskip .07in
%\centerline{(c)\hskip 3.5in (d)}

\caption{(a) Spike bursts of model (\ref{eqDengModel}) with the parameter values: $\Kabattery=-62$, $\Kbarconductance=0.015$, $\Kbconstant=16$, $\Naabattery=65$, $\Nabarconductance=70$, $\Nabconstant=20$,  $\Gabattery=-50$, $\Gbarconductance=15$, $\Gbconstant=10$, $C=1$, $\KTimeConstant=3$, $\NaGTimeConstant=80$,  $\Clabattery=-10$, $\Clbarconductance=10$, $\Clbconstant=50$, and $\ClTimeConstant=0.02$. The gating current and the chloride current are magnified four times for a better view. (b) Exactly the same parameter values for the 3-D model (\ref{eqDengModel3}). The $V$-nullcline shown is its intersection with $\Clpcurrent=0$. Dark dash line is the projected $\Clpcurrent$-nullcline with $\Clpcurrent=0$ and the light dash lines mark the nullcline's effective range for the spike burst.}\label{figSpikeBurst}
\end{figure}
%%%%%%%%%%%%%%%%%%%%%%%%%%%%%%%%%%%%%%%%%%%%%%%%%%%%%%%%%%%%%
%%%%%%%%%%%%%%%%%%%%%%%%%%%%%%%%%%%%%%%%%%%%%%%%%%%%%%%%%%%%%
%%%%%%%%%%%%%%%%%%%%%%%%%%%%%%%%%%%%%%%%%%%%%%%%%%%%%%%%%%%%%

But in order to sustain multiple bursts, another ion current is needed to drive the bursting-spiking orbit between the two regimes, being separated and redirected by the middle saddle-node point. The dynamics is into the spiking regime when an orbit is directed to the right of the point and into the quiescent regime when the orbit is directed to the left of the point. For illustration purpose, we will include the chloride ion Cl$^-$ current to the membrane model. We will assume it has a negative Nernst potential $\Clabattery<0$ above the resting membrane potential $\mbattery$, an exponential activation conductance just like the other two ion species with moderate intrinsic conductance $\Clbarconductance$ and a comparable activation voltage parameter $\Clbconstant$. Last, we will assume that it is current-adapted and slowly so $0<\ClTimeConstant\ll 1$. The resultant system is as follows
\begin{equation}\label{eqDengModel2}
\left\{\!\!\!\!\begin{array}{ll} & C\Vvoltageprime=-[\Kbarconductance n (\Vvoltage-\Kabattery)+\Nabarconductance m (\Vvoltage-\Naabattery)+\Gbarconductance h (\Vvoltage-\Gabattery)+\Clpcurrent-\Ecurrent]\\
& {n}' = \KTimeConstant(e^{(\Vvoltage-\Kabattery)/\Kbconstant}-n)\\
& {m}' = \NaGTimeConstant(e^{(\Vvoltage-\Naabattery)/\Nabconstant}-m)\\
& {h}' = \NaGTimeConstant(e^{-(\Vvoltage-\Gabattery)/\Gbconstant}-h)\\
& {\Clpcurrent}' = \ClTimeConstant(\Clbarconductance e^{(\Vvoltage-\Clabattery)/\Clbconstant} (\Vvoltage-\Clabattery)-\Clpcurrent)\\
\end{array}\right.
\end{equation}
We also note that for many types of neuron and excitable membranes their spike-burst dynamics are determined by sodium, potassium, and calcium ions (Ca$^{++}$) with the role of the sodium in the model above being replaced by the calcium (whose Nernst potential is usually higher still than that of the sodium ion) and the role of the chloride replaced by the sodium so that the region of spikes is always above the zero membrane potential $V=0$. For mathematical mechanisms which generate spike bursts, the prototypical model above is nonetheless rather typical and illuminating. More specifically, Fig.\ref{figSpikeBurst}(a) shows a simulation of such a spike burst. One can see that gating again plays an important role for this behaviour, shutting down the burst when it is activated and permitting spiking when it recedes. Also, it shows that the addition of the slow chloride current increases during the spiking phase until sending the dynamics into the quiescent phase and decreases during the silent phase until sending the dynamics into the active phase.

Spike burst is a 3-dimensional phenomenon just like action potential is basically a 2-dimensional one. To demonstrate this point, let assume the sodium and gating dynamics is the fastest with $\NaGTimeConstant\sim\infty$ so that their conductances equilibriumize instantaneously with $m=e^{(V-\Naabattery)/\Nabconstant}$ and $h=e^{(V-\Gabattery)/\Gbconstant}$. This eliminates two differential equations from (\ref{eqDengModel2}) above to obtain the 3-dimensional system below
\begin{equation}\label{eqDengModel3}
\left\{\!\!\!\!\begin{array}{ll} & C\Vvoltageprime=-[\Kbarconductance n (\Vvoltage-\Kabattery)+\Nabarconductance e^{(\Vvoltage-\Naabattery)/\Nabconstant} (\Vvoltage-\Naabattery)+\\
&\qquad\qquad\qquad \Gbarconductance e^{-(\Vvoltage-\Gabattery)/\Gbconstant} (\Vvoltage-\Gabattery)+\Clpcurrent-\Ecurrent]\\
& {n}' = \KTimeConstant(e^{(\Vvoltage-\Kabattery)/\Kbconstant}-n)\\
& {\Clpcurrent}' = \ClTimeConstant(\Clbarconductance e^{(\Vvoltage-\Clabattery)/\Clbconstant} (\Vvoltage-\Clabattery)-\Clpcurrent)\\
\end{array}\right.
\end{equation}
Fig.\ref{figSpikeBurst}(b) is the phase space of this system at the cross-section $\Clpcurrent=0$. The $V$-nullcline and the $n$-nullcline clearly show the all-or-noting firing configuration, having three intercept points. It also shows the spikes lie to the right side of the middle saddle-node of the $Vn$-subsystem. As most part of the spikes lies to the right side of $\Clpcurrent$-nullcline in which $\Clpcurrent'>0$, $\Clpcurrent$ increases to become more positive. The outward current inhibits spiking, lowering the $V$-nullcline until the orbit is caught to the left side of the middle saddle-node point and the spikes are switched off. But the quiescent phase lies to the left side of the $\Clpcurrent$-nullcline in which $\Clpcurrent'<0$, $\Clpcurrent$ decreases to become more negative, changing it into an excitatory current to drive the membrane into its spiking regime again. This is the basic 3-dimensional blueprint for continuous bursts of spikes in all higher dimensions. For exactly the same parameter values, the lower dimensional system (\ref{eqDengModel3}) and the higher dimensional counterpart (\ref{eqDengModel2}) are all comparable except for possibly a different number of spikes. That is, if we project the higher dimensional bursts of spikes from Fig.\ref{figSpikeBurst}(a) to its $Vn$-space, it should look like the phase portrait Fig.\ref{figSpikeBurst}(b) of the lower dimensional counterpart.

\bigskip
\noindent{\bf 4. Concluding Remark.} Our mathematical modeling of the squid axon began with a model for the sodium-potassium ion exchanger pump which led to a justification of the Nernst potentials for both ion species and to the modeling of the ion currents by battery-driven conductors. Following upon the voltage-dependent conductance finding of Hodgkin and Huxley we introduced the exponential activation model for both opening and closing of ion channels. This activation model unifies two seemingly different types of activation in potassium and sodium channels. We also introduced the exponential voltage-gating model for the sodium channel which together with the sodium activation model automatically led to $N$-shaped sodium-gating characteristics, solving the negative conductivity problem which has puzzled researchers of many generations. By incorporating Hodgkin and Huxley's time adaptation for channel conductances, our model is capable of generating the all-or-nothing action potentials as a must-be consequence to voltage-gating as well as spike bursts with the all-or-nothing firing configuration. These voltage-gating related properties have never been demonstrated in any existing model.

More specifically, our model enables the following narrative on action potential generation in mechanistic details unobtainable from the HH model. In the absence of external excitation the membrane is kept at rest by voltage-gating. When depolarizing excitation exceeds a threshold above the maximal voltage-gated current, the voltage-gating current drops exponentially to zero, allowing the membrane to depolarize which in turn opens up the sodium gates. The increased inward sodium current further depolarizes the membrane ($dV/dt>0$) which in turn opens up not only more sodium gates but also potassium gates as well in the manner that their channel conductances grow exponentially with the depolarizing voltage. However, the inward sodium current must slow down as it is buffered by the potential difference between the membrane potential and its positive Nernst potential ($\Napcurrent=\Naconductance(V)(V-\Naabattery)$), but in contrast the outward potassium current runs away exponentially from its negative Nernst potential. The outward potassium current must eventually catch up to the inward sodium current in magnitude so that the net ion and gating current becomes outward and the direction of the across-membrane voltage is reversed ($dV/dt\le 0$). This reversal activates the closing of both ion gates, closing up them exponentially fast. This in turn speeds up the downfall of the membrane potential. The hyperpolarizing potential may not pass its resting potential if the outward potassium current closes itself too quickly, giving rising to continuous spiking of some sort. Otherwise, the across-membrane potential must overshoot the voltage-gating potential ($V<\Gabattery$), in which case the already-activated voltage-gating conductance is near its intrinsic conductance. From this point on the system must converge to the membrane resting potential whether or not the voltage overshoots it. This is because of two interplays of the currents. One, the voltage-gating current becomes negative below its reversal potential. Two, the sodium and potassium currents become so much smaller than the small inward voltage-gating current. So the latter becomes dominating, effectively halting membrane hyperpolarization and then reversing the direction of the hyperpolarizing membrane potential ($dV/dt\ge 0$), and bringing the membrane to its re-polarized resting state yet again.

One conclusion seems obvious that our model (\ref{eqDengModel}) should replace the HH model as the basic template for excitable membranes for future researches in theoretical neuroscience. Another takeaway from our result is that however complicated molecular biological processes may be it is not only possible but also imperative to model them mechanistically. Phenomenological models can fit but only mechanistic models can both fit and explain. Mathematical modeling is to find the mechanistic equation to which nature fits as a solution, as Newton demonstrated centuries ago. Richard Feynman believed that if we cannot lecture a theory to sophomore physics students then we must not understand the theory. Our model should pass his test.

%\bigskip
%\noindent {\bf Acknowledgement:}
%\begin{thebibliography}{99}
%\bibitem{Cric68}
%\end{thebibliography}

\bibliography{bibcircuit15}
\bibliographystyle{ieeetr}

\end{document}

%% file: definitions.tex
\newcommand{\Kpcurrent}{I_{\rm K}}
\newcommand{\Napcurrent}{I_{\rm Na}}
\newcommand{\Lpcurrent}{I_{\rm L}}
\newcommand{\Gpcurrent}{I_{\rm G}}

\newcommand{\Clpcurrent}{I_{\rm Cl}}

\newcommand{\NaTimeConstant}{\tau_{ _{\rm Na}}}
\newcommand{\KTimeConstant}{\tau_{ _{\rm K}}}

\newcommand{\GTimeConstant}{\tau_{ _{\rm G}}}
\newcommand{\NaGTimeConstant}{\tau_{ _{\rm NaG}}}

\newcommand{\ClTimeConstant}{\tau_{ _{\rm Cl}}}

\newcommand{\NaIVf}{f_{ _{\rm Na}}}
\newcommand{\KIVf}{f_{ _{\rm K}}}
\newcommand{\LIVf}{f_{ _{\rm L}}}
\newcommand{\GIVf}{f_{ _{\rm G}}}

\newcommand{\Ecurrent}{I_{\rm ext}}

\newcommand{\Kabattery}{E_{\rm K}}
\newcommand{\Naabattery}{E_{\rm Na}}
\newcommand{\Clabattery}{E_{\rm Cl}}

\newcommand{\mbattery}{E_{\rm m}}
\newcommand{\Gabattery}{E_{\rm G}}

\newcommand{\Kconductance}{g_{ _{\rm K}}}
\newcommand{\Naconductance}{g_{ _{\rm Na}}}

\newcommand{\Gconductance}{g_{ _{\rm G}}}

\newcommand{\Kbconstant}{b_{ _{\rm K}}}
\newcommand{\Nabconstant}{b_{ _{\rm Na}}}
\newcommand{\Gbconstant}{b_{ _{\rm G}}}

\newcommand{\Clbconstant}{b_{ _{\rm Cl}}}

\newcommand{\Kbarconductance}{{\bar g}_{ _{\rm K}}}
\newcommand{\Nabarconductance}{{\bar g}_{ _{\rm Na}}}

\newcommand{\Clbarconductance}{{\bar g}_{ _{\rm Cl}}}

\newcommand{\Gbarconductance}{{\bar g}_{ _{\rm G}}}

\newcommand{\Vvoltage}{V}
\newcommand{\Vvoltageprime}{{V}'}